%% file: pulse_manu.tex
\documentclass[a4paper,11pt]{article}
\pdfoutput=1 

\usepackage{jinstpub} 
\usepackage{lineno,hyperref}
\modulolinenumbers[1]
\usepackage{graphicx}
\usepackage{float}
\usepackage{threeparttable}
\usepackage{subfigure}
\usepackage{amsmath,amsfonts,amsthm} 
\usepackage{bm}
\usepackage[load-configurations=abbreviations,detect-all=true]{siunitx}

\title{\boldmath Timing and characterization of shaped pulses with MHz ADCs in a detector system: a comparative study and deep learning approach}

\author[a]{Pengcheng Ai,}
\author[a,1]{Dong Wang,\note{Corresponding author.}}
\author[a]{Guangming Huang,}
\author[a]{Ni Fang,}
\author[a]{Deli Xu,}
\author[b]{Fan Zhang}

\affiliation[a]{Central China Normal University, No.152 Luoyu Road, Wuhan, Hubei, 430079 P.R.China}
\affiliation[b]{Hubei University of Technology, No.28 Nanli Road, Wuhan, Hubei, 430068 P.R.China}

\emailAdd{dongwang@mail.ccnu.edu.cn}

\abstract{Timing systems based on Analog-to-Digital Converters are widely used in the design of previous high energy physics detectors. In this paper, we propose a new method based on deep learning to extract the time information from a finite set of ADC samples. Firstly, a quantitative analysis of the traditional curve fitting method regarding three kinds of variations (long-term drift, short-term change and random noise) is presented with simulation illustrations. Next, a comparative study between curve fitting and the neural networks is made to demonstrate the potential of deep learning in this problem. Simulations show that the dedicated network architecture can greatly suppress the noise RMS and improve timing resolution in non-ideal conditions. Finally, experiments are performed with the ALICE PHOS FEE card. The performance of our method is more than 20\% better than curve fitting in the experimental condition.}


\arxivnumber{} 

\begin{document}	
\maketitle
\flushbottom

\section{Introduction}

Pulse timing is a common problem in high energy physics \cite{grzywacz2003applications}, optics \cite{samain1998timing}, telecommunication \cite{han2017timing} and many other applied physics disciplines. Among feasible methods, fast electronic readout systems provide a cost-effective and robust solution with relatively high timing resolution. In many engineering circumstances, we care more about the availability and practicality than technical indicators. Electronic timing systems are usually good candidates for these applications.

In high energy physics, accurate timing, along with energy and position information, is needed to reconstruct the collision events so as to discriminate against backgrounds \cite{alice2014performance} and identify phenomenons of interest \cite{aad2012observation}. Several kinds of detectors can provide the time information. For example, Time-of-Flight detectors can measure the time of incoming events directly; Time Projection Chambers (TPC) and calorimeters can measure the pulse signal and infer the time afterwards; silicon detectors and pixel sensors can measure the hit information and offer an auxiliary time stamp, and so on. The final reconstructed event is a combination and coincidence of multiple sources of detectors.

There are two major branches of timing systems: systems based on Analog-to-Digital Converters (ADC) and systems based on Time-to-Digital Converters (TDC). In general, TDC-based systems are specialized in time measurement and can achieve a precision of tens of picoseconds \cite{antonioli200320} when configured properly. In spite of their high precision, the major drawback of TDC-based systems is that they lack the necessary amplitude information which is critical in some applications. If both time and amplitude are of interest, ADC-based systems are good alternatives to TDC-based systems. The empirical timing precision for ADC-based systems is in the order of nanoseconds.

For ADC-based systems, a typical work flow can be described as follows. The original signal from TPCs or calorimeters is preprocessed by Charge Sensitive pre-Amplifiers (CSA) to get a step-like signal. Afterwards, this signal is fed to Front-End Electronics (FEE). The signal conditioning on the FEE board includes buffering, amplifying and bandpass filtering by CR-RCn shapers. Finally, the signal is sampled by ADCs with the prescribed precision and data depth. The recorded ADC samples can serve multiple purposes. For a classification task, the shaped pulse signal can be used to discriminate between particles or physical events \cite{mauri2018pulse, mahata2018particle, akerib2018liquid, ashida2018separation}. For a regression task, timing or other pulse information is extracted from the digitized pulse signal \cite{kaspar2017design}.

To obtain the time from a finite set of ADC samples, we can use an estimated fitting function and perform curve fitting to get estimated values of underlying parameters. Curve fitting is a standard inference method in the time domain and it shows promising properties under certain conditions (See section \ref{sec:adv-curve-fitting}). However, its applicability and accuracy rely on the fitting function and the ideal form of noise heavily. As a result, the actual performance of curve fitting is limited by experimental conditions of ADC-based systems \cite{fish2017electronic}.

Recently, \emph{deep learning} \cite{lecun2015deep} as a renewed machine learning technique has progressed rapidly. It has been successfully used for particle/event discrimination and identification at the pulse level \cite{griffiths2018pulse}, the pixel level \cite{adams2018deep} and the voxel (three-dimensional) level \cite{ai2018three}. In view of the fact that neural networks are applicable to classification tasks as well as regression tasks, it is meaningful to explore the capability of deep learning in the above-mentioned pulse timing problem.

In this paper, we mainly discuss the deep learning approach to pulse timing based on a comparison between curve fitting and the proposed method. Section \ref{sec:alice-phos} briefly introduces the project background and the mathematical form of the researched pulse. Section \ref{sec:curve-fitting} explains the traditional curve fitting method by theoretical analysis and simulation studies. Section \ref{sec:deep-learning} gives a comparative study and the details of the new approach of deep learning. Section \ref{sec:exp-results} discusses the experiments we conduct and shows the experimental results. Finally, a conclusion is drawn in section \ref{sec:conclusion}.

\section{ALICE PHOS electronics}
\label{sec:alice-phos}

The ALICE PHOS detectors \cite{muller2006config} refer to the Photon Spectrometers designed for the ALICE experiment \cite{aamodt2008alice}. The detectors were produced in 2007 and scheduled for the first p+p collisions at LHC in 2008 \cite{torii2009phos}. The scintillator is made of lead tungstate crystals and mainly used to detect high energy photons (up to 80 GeV). An Avalanche Photo-Diode (APD) receives the scintillation and converts it to an electrical signal, which is applied to a CSA near the APD. The output of the CSA is connected to the FEE card via a flat cable.

The FEE card has 32 independent readout channels, each of which is connected to two shaper sections with high gain and low gain. The CR-RC2 signal shapers are made up of discrete components on a 12-layer Printed Circuit Board (PCB). For each channel, there are two overlapping 10-bit ADCs at the terminations of the two shapers, which give an equivalent dynamic range of 14 bits. The sampling rate of the ADCs is fixed to 10 MS/s. The same readout plan and PCB layout were adopted by ALICE EMCal detectors \cite{fantoni2011emcal}, which refer to the ALICE Electromagnetic Calorimeters. The major difference of FEE cards between PHOS and EMCal lies in the shaping time of the shapers. For PHOS, the designate shaping time is 1 $\mu$s; however for EMCal, we use different resistors and capacitors to achieve a shaping time of 100 ns.

The CR-RC2 shaper is a bandpass filter in the frequency domain. In the time domain, its response to an ideal step signal can be formulated as the equation below:

\begin{equation} \label{equ:pulse}
	f(t) = \begin{cases} K \left(\frac{t-t_0}{\tau_p}\right)^2 \cdot e^{-2 \cdot (\frac{t-t_0}{\tau_p})} + b, &\text{for } t \geq t_0 \\
	b &\text{for } t < t_0 \end{cases}
\end{equation}

\noindent where $t_0$ is the start time, and $b$ is the pedestal. $K$ is originally defined as $\frac{2Q \cdot A^2}{C_f}$ which is a variable related to the energy of the incoming photon, where $Q$ is the APD charge, $A$ is the shaper gain and $C_f$ is charging capacitance of the CSA. In our simulations, without changing the nature of the problem, we use $K$ as a normalization factor for numerical purposes. $\tau_p$ is the peaking time defined as the interval between the start of the semi-Gaussian pulse and the moment when $f(t)$ reaches its maximum value. The relation between the shaping time $\tau_0$ and the peaking time $\tau_p$ is $\tau_p = n \cdot \tau_0$. For the CR-RC2 shaper structure, $n$ equals 2, so the peaking times for the PHOS and EMCal are 2 $\mu$s and 200 ns, respectively.

Since the CR-RCn shaper is representative for most applications in high energy physics, in the latter sections we center on the pulse function in equation \ref{equ:pulse} to discuss different timing methods.

\section{Curve fitting method}
\label{sec:curve-fitting}

Curve fitting is a traditional model fitting technique mainly aimed at finding the parameterized mathematical relations between two or more variables. Classical linear curve fitting can be directly solved by the least squares method, and nonlinear curve fitting can be solved by the trust region and Levenberg-Marquardt methods \cite{marquardt1963algorithm}. In the pulse timing scenario, the main purpose of curve fitting is to determine the desired parameters related to the time information. In the following subsections, we analyze the curve fitting method in terms of its capability to reveal the ground-truth parameters under various conditions.

\subsection{Theoretical analysis}

\subsubsection{Summary and notations}

We consider the following nonlinear least squares problem:

\begin{equation}
\begin{aligned}
	\text{minimize } & S \\
	= \text{minimize } & \sum_{i=1}^n r_i^2 \\
	= \text{minimize } & \sum_{i=1}^n \left[y_i - f(t_i; \bm{\beta}, \bm{\theta})  \right]^2
\end{aligned}
\end{equation}

\noindent where $S$ is the sum of squared residuals to minimize, $r_i$ is the i-th residual, $y_i$ is the i-th observed value (from ADC), and $t_i$ is the i-th time value. There is some noise residing in the observed value $y_i$, and we denote this noise term as $n_i$. Besides, $\bm{\beta}$ is the \emph{fitting parameters} and $\bm{\theta}$ is the \emph{system parameters}. The division of fitting parameters and system parameters is made according to our understanding of the problem and practical issues. It is not recommended to set two parameters with high correlation as fitting parameters at the same time, which will cause instability to the fitting process.

It should be noted that the above formulation is a general framework for the fitting problem. Usually we choose a function family $f(t; \bm{\beta}, \bm{\theta_m})$ for curve fitting. However, $f(t; \bm{\beta}, \bm{\theta_m})$ is only a subset of the underlying possible functions $f(t; \bm{\beta}, \bm{\theta})$. We denote the reference fitting function as $f(t; \bm{\beta_0}, \bm{\theta_0})$ in section \ref{sec:adv-curve-fitting} and section \ref{sec:quan-analysis}.

\subsubsection{The advantage of curve fitting in the ideal condition}
\label{sec:adv-curve-fitting}

In this part, we assume that the selected fitting function is accurate (i.e. $\bm{\theta}$ is fixed to $\bm{\theta_0}$ and $\bm{\theta_m} = \bm{\theta_0}$), and the noise distribution is strictly Gaussian with a fixed variance $\sigma$. Under these assumptions, the distribution of the observed value can be written as:

\begin{equation}
	y_i = f(t_i; \bm{\beta_0}, \bm{\theta_0}) + n_i \sim N\left( f(t_i; \bm{\beta_0}, \bm{\theta_0}), \sigma^2 \right)
\end{equation}

Since the Gaussian distribution is $P(x|\mu, \sigma^2) = \frac{1}{{\sqrt{2\pi}\sigma}}e^{{{ - \left( {x - \mu } \right)^2 } \mathord{\left/ {\vphantom {{ - \left( {x - \mu } \right)^2 } {2\sigma ^2 }}} \right. \kern-\nulldelimiterspace} {2\sigma ^2 }}}$, the corresponding log-likelihood function is:

\begin{equation} \label{equ:log-likelihood}
\begin{aligned}
	L(y_1, y_2, \ldots, y_n; \bm{\beta_0}, \bm{\theta_0}) & = \ln\prod_{i=1}^n P(y_i|f(t_i; \bm{\beta_0}, \bm{\theta_0}), \sigma^2) \\
	                         & = -\frac{1}{2\sigma^2}\sum_{i=1}^n\left[y_i-f(t_i; \bm{\beta_0}, \bm{\theta_0})\right]^2 + const
\end{aligned}
\end{equation}

The equation \ref{equ:log-likelihood} implies that, in the ideal condition, using curve fitting to minimize the sum of squared residuals $S$ is equivalent to maximizing the log-likelihood function of the noise distributions. In other words, curve fitting gives the \emph{maximum likelihood estimators} of fitting parameters. This claim reveals the statistical properties of the curve fitting method. It is based on a hypothesis of Gaussian noise distributions, which is a useful prior when our knowledge about the system is limited.  

\subsubsection{Quantitative analysis of drift, change and noise}
\label{sec:quan-analysis}

In reality, the assumptions in section \ref{sec:adv-curve-fitting} are usually not valid. Variations in the fitting function and the noise make the problem much more complicated. In this paper, we consider three types of variations which are representative in high energy physics:

\begin{enumerate}
	\item \emph{Long-term drift}. This kind of variation refers to the deviation in the system parameters $\bm{\theta}$ after the circuit board is fabricated. It can also represent the persistent change between two calibration runs. It will affect the pulse function consistently so that the event-by-event characteristics stay the same for ADC sampling values.
	
	\item \emph{Short-term change}. This kind of variation refers to the deviation in the system parameters $\bm{\theta}$ between two events. It will change according to the current status of the detector, but its effect is near-identical to all ADC sampling values in a single event. In other words, the event-by-event characteristics will change in the operation of the experiment.
	
	\item \emph{Random noise}. This kind of variation refers to the randomized noise $n_i$ residing in the observed value $y_i$. It will vary between ADC samples in a single event. Since it is random, the actual value of the noise is not predictable. However, its statistical features can be determined in advance.
\end{enumerate}

Next, we will introduce these variations into the curve fitting. We only consider the variations that are near the reference point so that the fitting result will not be rejected by the fitting process (i.e. without increasing the chi-square criterion significantly). When the above variations are present, by using the first-order approximation we can formulate $y_i$ as:

\begin{equation} \label{equ:deviation-y}
	y_i = f(t_i; \bm{\beta_0}, \bm{\theta_0}) + \sum_j \frac{\partial f(t_i; \bm{\beta_0}, \bm{\theta_0})}{\partial \theta_j}\Delta\theta_j + n_i
\end{equation}

Since we use the reference system parameters in the curve fitting, non-ideal $y_i$ will cause a change in the fitting parameters. By using the first-order approximation:

\begin{equation} \label{equ:deviation-ft}
	f(t_i; \bm{\beta}, \bm{\theta_0}) = f(t_i; \bm{\beta_0}, \bm{\theta_0}) + \sum_j \frac{\partial f(t_i; \bm{\beta_0}, \bm{\theta_0})}{\partial \beta_j}\Delta\beta_j
\end{equation}

Curve fitting tries to minimize the sum of squared residuals by varying $\bm{\beta}$. By applying the first-order necessary condition for a minimum, we get the following equation:

\begin{equation} \label{equ:gradient_S}
	\nabla_{\bm{\beta}}S = \nabla_{\bm{\beta}}\sum_{i=1}^n r_i^2 = \nabla_{\bm{\beta}}\left[ \sum_{i=1}^n \left(y_i - f(t_i; \bm{\beta}, \bm{\theta_0}) \right)^2 \right] = 0
\end{equation}

By substituting equation \ref{equ:deviation-y} and equation \ref{equ:deviation-ft} into equation \ref{equ:gradient_S} and solving the system of linear equations, we can get the following expression:

\begin{gather}
	(\bm{J^T J})\bm{\Delta \beta} = \bm{J^T} (\bm{P \Delta\theta} + \bm{n}) \\
	\text{where  } J_{ij} = \frac{\partial f(t_i; \bm{\beta_0}, \bm{\theta_0})}{\partial \beta_j} \quad P_{ij} = \frac{\partial f(t_i; \bm{\beta_0}, \bm{\theta_0})}{\partial \theta_j} \nonumber
\end{gather}

If $\bm{J^T J}$ is nonsingular, the deviation in the fitting parameters can be solved by:

\begin{equation} \label{equ:dev-solve}
	\bm{\Delta \beta} = (\bm{J^T J})^{-1} \bm{J^T} (\bm{P \Delta\theta} + \bm{n})
\end{equation}

In general, equation \ref{equ:dev-solve} is a generalization of linear curve fitting to nonlinear cases. It implies that, under first-order approximations, the deviation of the fitting parameters around the reference point is linearly dependent on the deviation of the system parameters and random noise.

\subsection{Simulation studies}
\label{sec:curve-fitting-sim}

To demonstrate the accuracy of first-order approximations in our pulse function, we compare the results from calculating equation \ref{equ:dev-solve} to the results from directly applying curve fitting. For the pulse function in equation \ref{equ:pulse}, we divide parameters in the following way without inducing a complicated function family:

\begin{equation}
\bm{\beta} = \{K, t_0 \}, \quad \bm{\theta} = \{\tau_p, b\}
\end{equation}

In the following simulations, we choose $K = 5.12$, $t_0 = 0.0$, $\tau_p = 2.0$, $b = 0.1$ as the reference point. The pulse is sampled from $t = 0.0$ to $t = 3.2$ with a period of $0.1$, so there are a total of 33 points. The value of $K$ ensures that the amplitude is renormalized to a range in the interior of $(0, 1)$. This parameterization is in accord with the PHOS electronics with 1 $\mu$s shaping time (section \ref{sec:exp-1us}).

\begin{figure}[htbp]
	\centering
	\subfigure[$K$ vs. $\tau_p$]{			
		\includegraphics[width=0.48\textwidth]{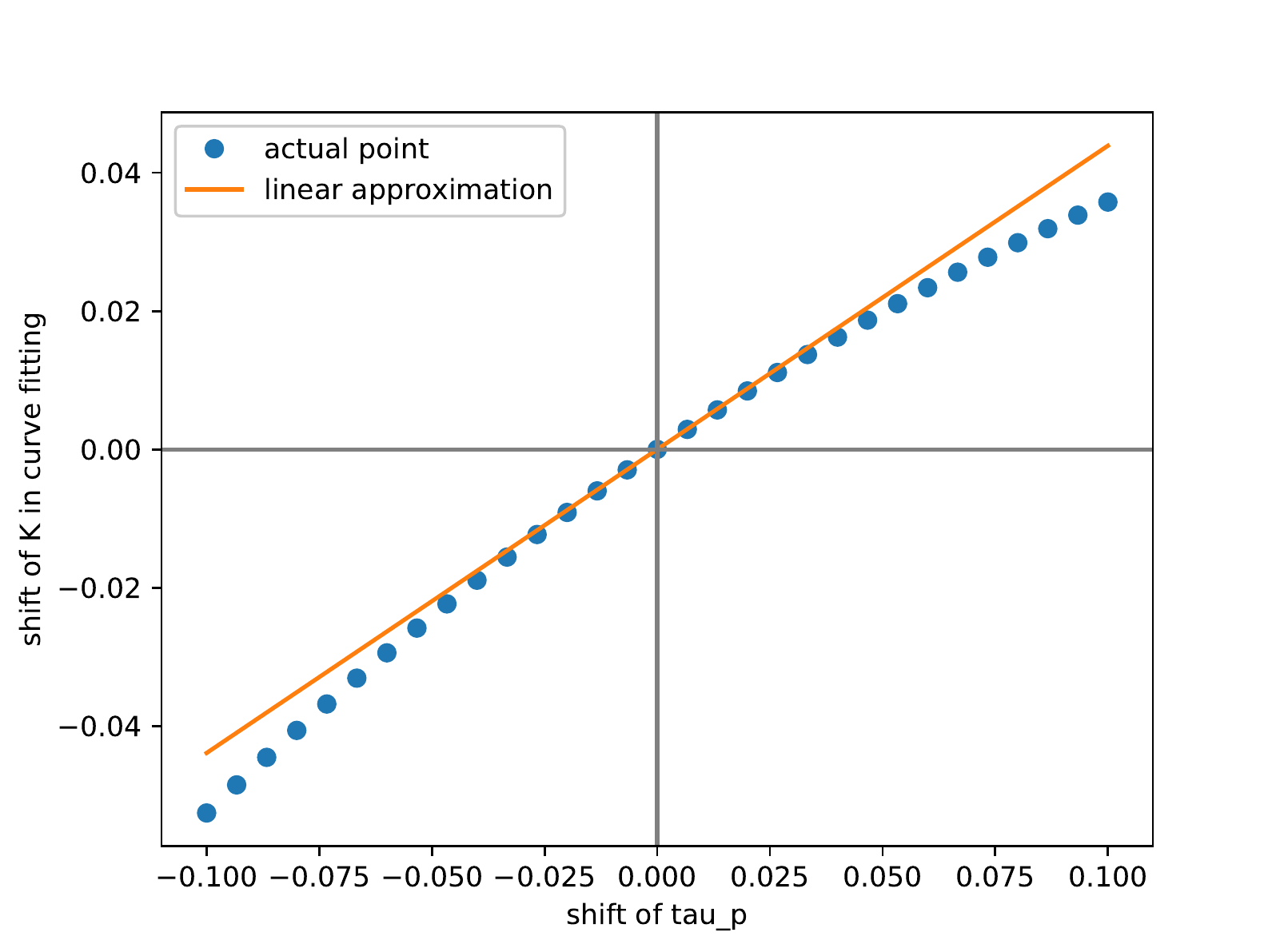}}		
	\subfigure[$t_0$ vs. $\tau_p$]{
		\includegraphics[width=0.48\textwidth]{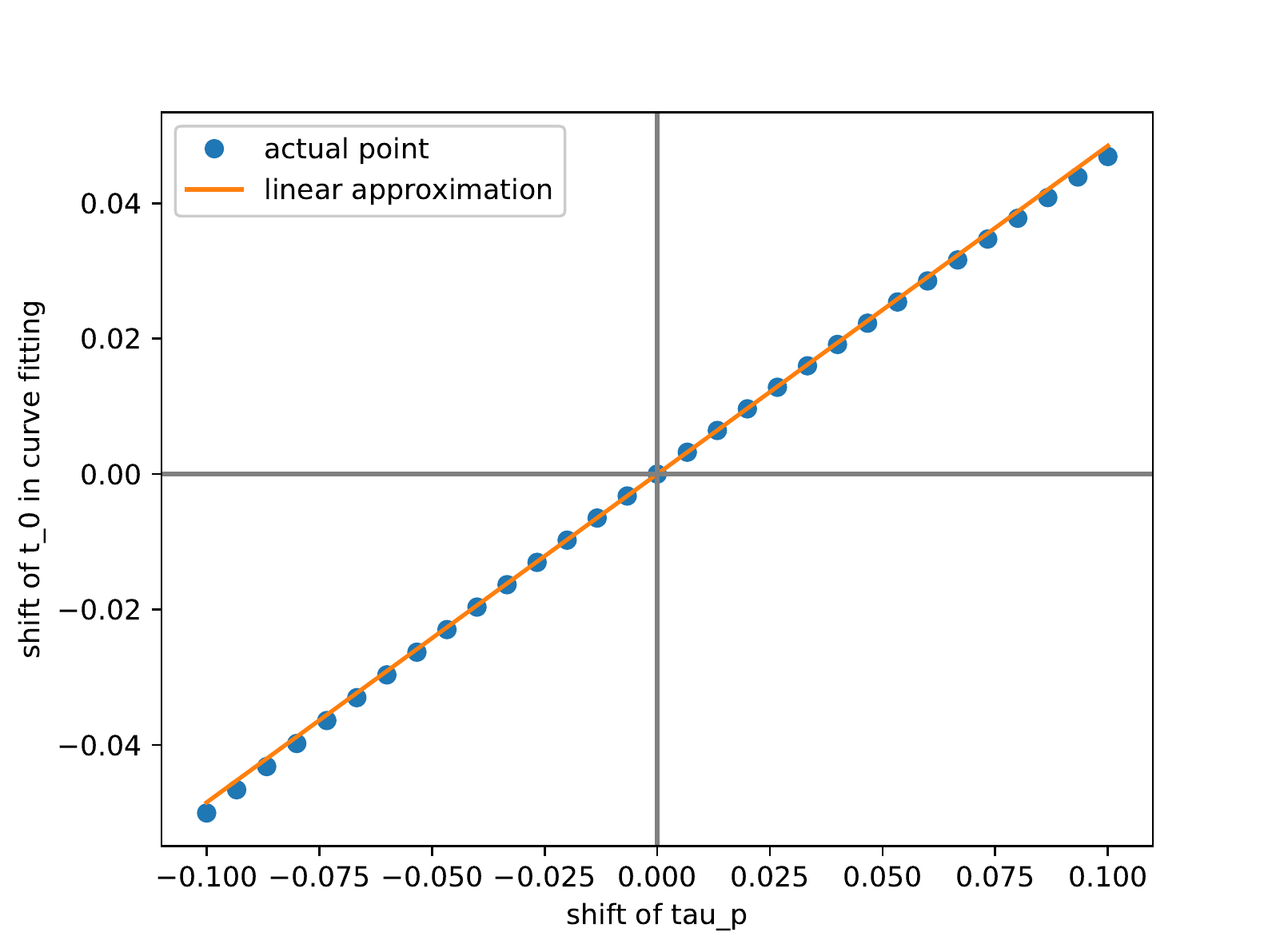}}
	\subfigure[$K$ vs. $b$]{			
		\includegraphics[width=0.48\textwidth]{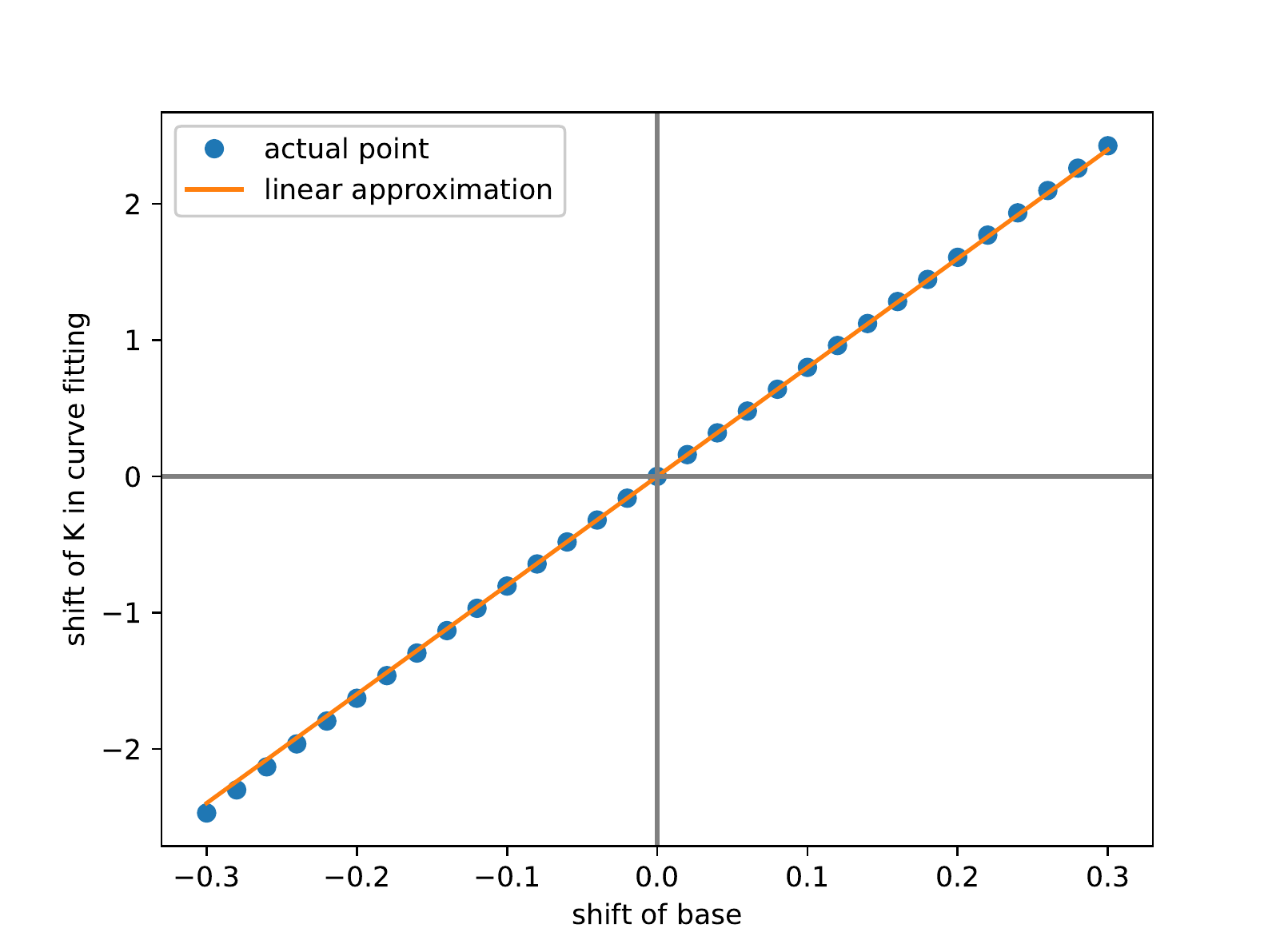}}		
	\subfigure[$t_0$ vs. $b$]{
		\includegraphics[width=0.48\textwidth]{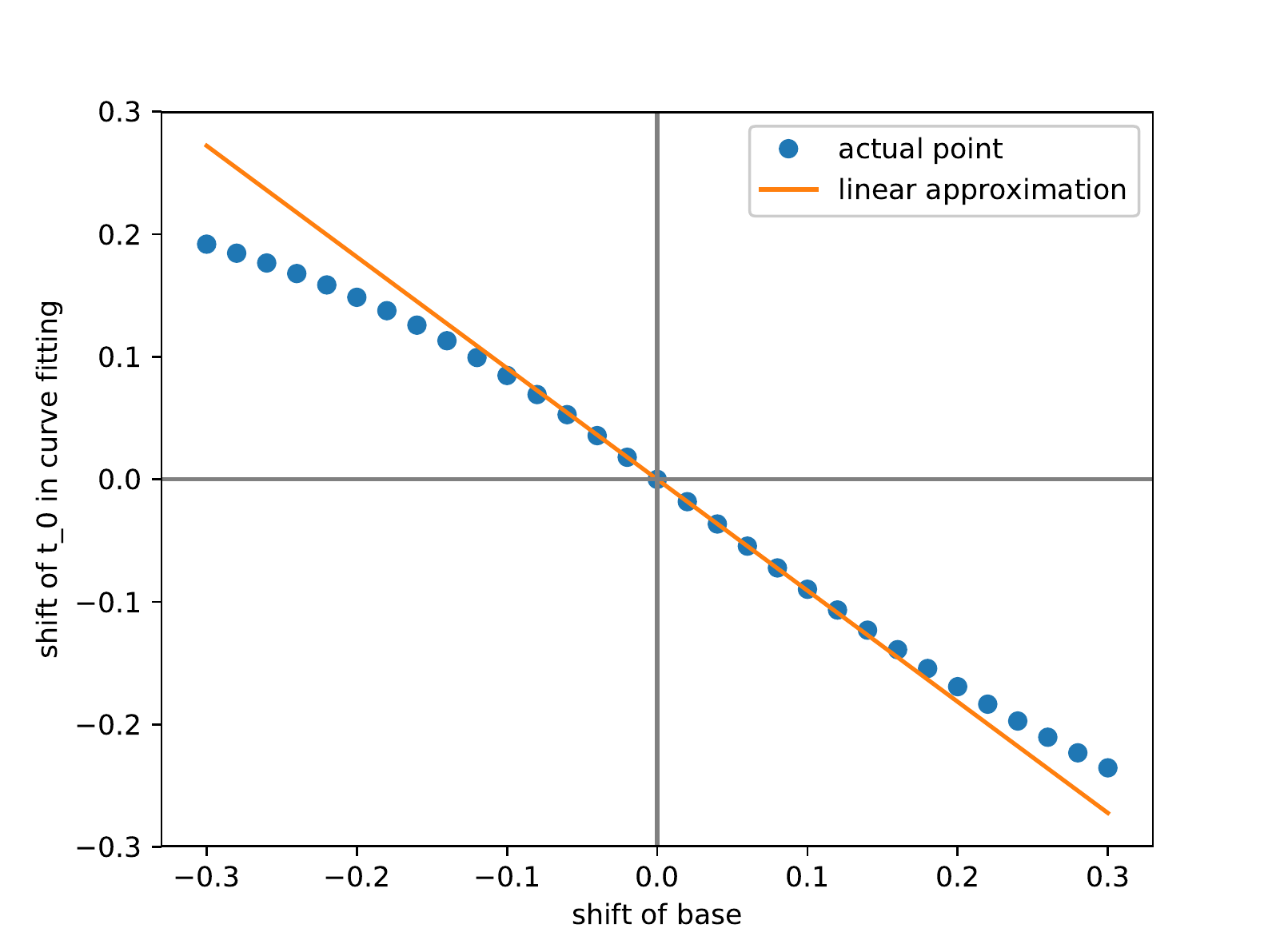}}
	\caption{\label{fig:curve-fitting-drift-change} A gathering of figures for drift and change simulations. Each figure compares the result from first-order (linear) approximations and the result from curve fitting directly.}
\end{figure}

\paragraph{Long-term drift and short-term change} These two kinds of variations are associated with system parameters $\bm{\theta}$. We separate $\tau_p$ and $b$ and study their influence on fitting parameters $K$ and $t_0$ respectively. The simulation results are shown in figure \ref{fig:curve-fitting-drift-change}. The solid line is calculated from first-order approximations, and the solid dots are generated from curve fitting. It can be seen that in a region near the reference point the first-order approximations are fairly accurate. This is especially true for ($t_0$, $\tau_p$) and ($K$, $b$) pairs, which have high correlations. In other two pairs, the discrepancy of first-order approximations and curve fitting is determined by higher order effects.

\begin{figure}[htbp]
	\centering
	\subfigure[$K$ vs. shift of clipped crystal ball]{			
		\includegraphics[width=0.48\textwidth]{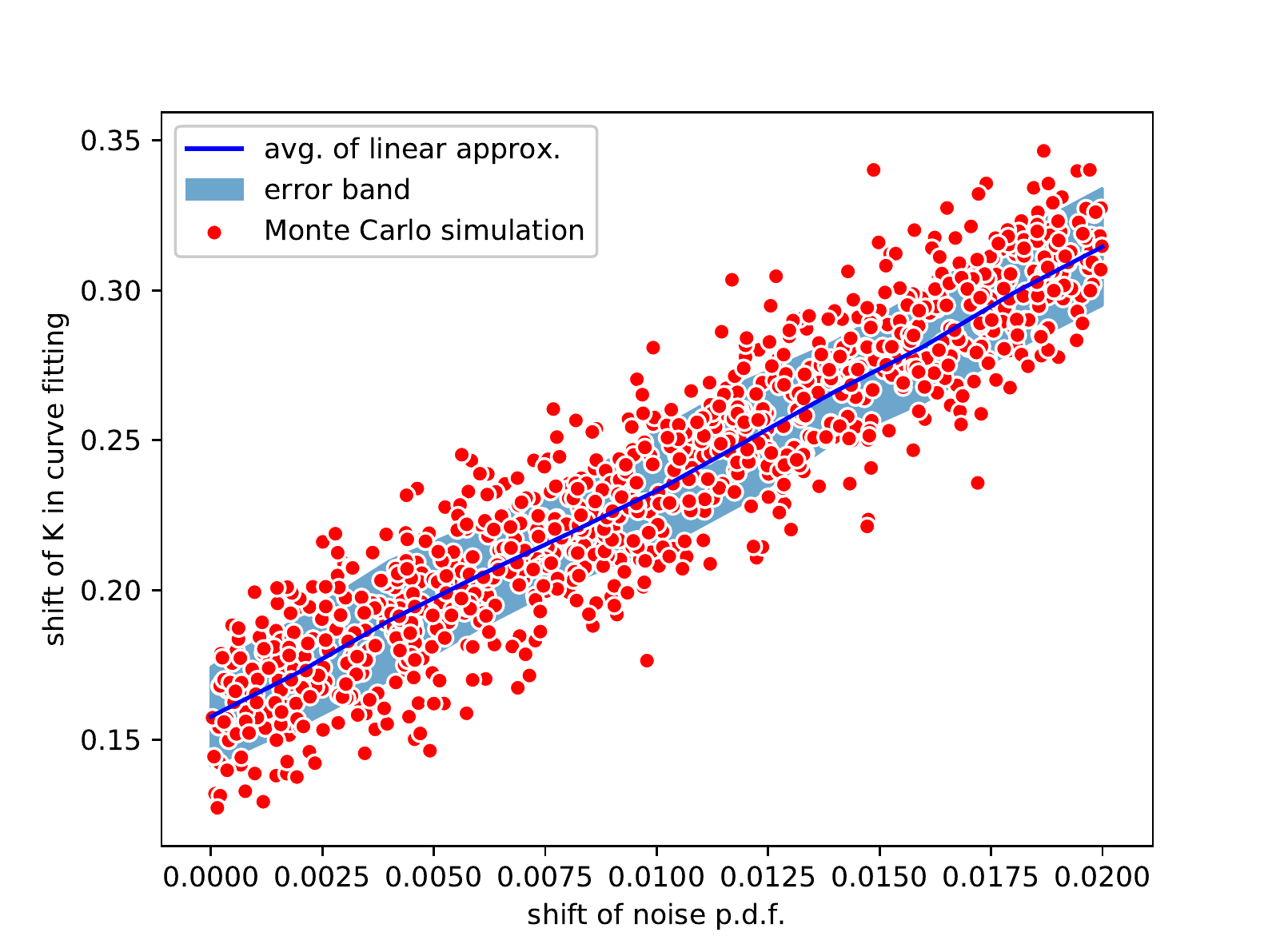}}		
	\subfigure[$t_0$ vs. shift of clipped crystal ball]{
		\includegraphics[width=0.48\textwidth]{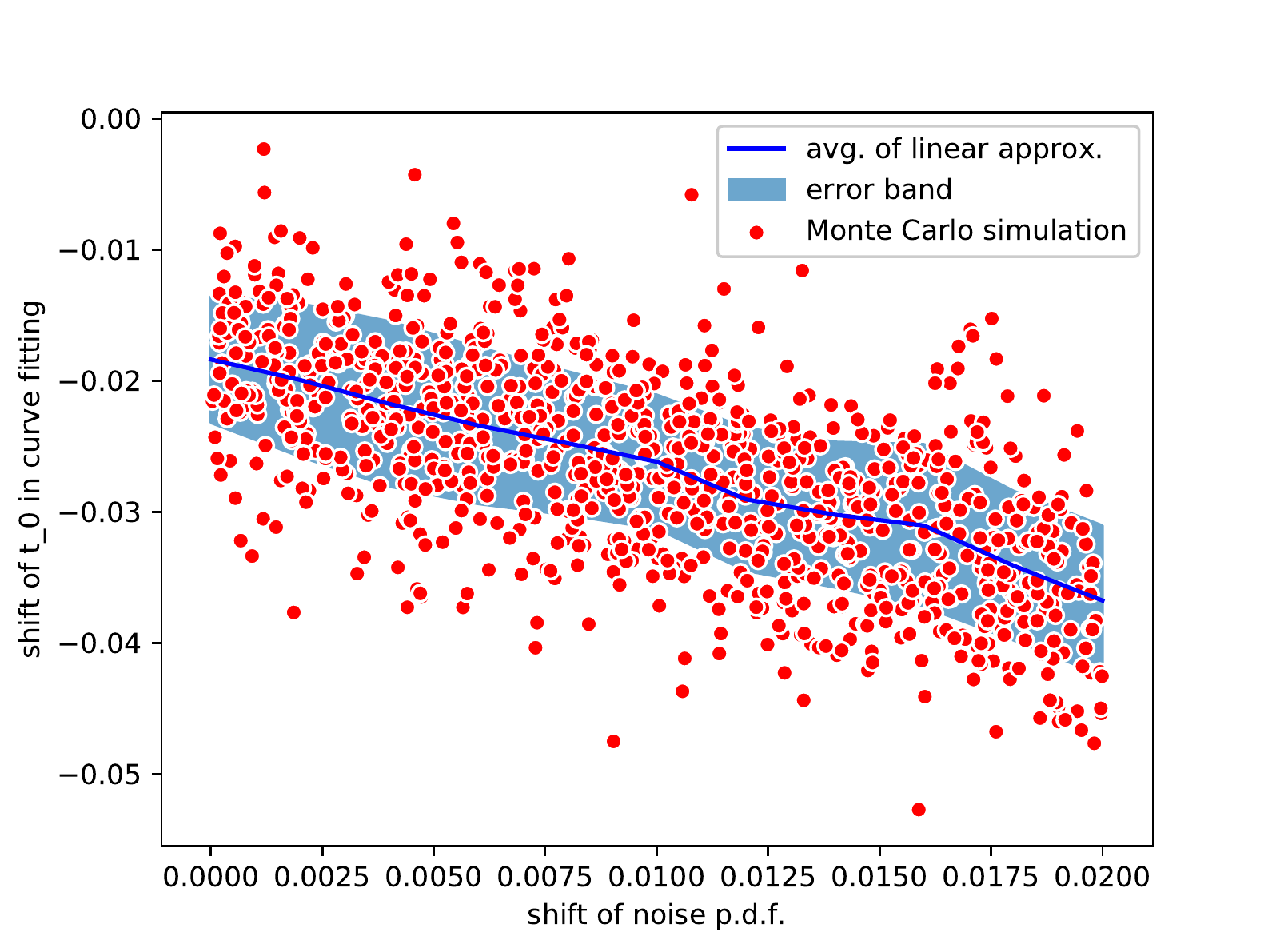}}
	\subfigure[$K$ vs. shift of clipped Moyal]{			
		\includegraphics[width=0.48\textwidth]{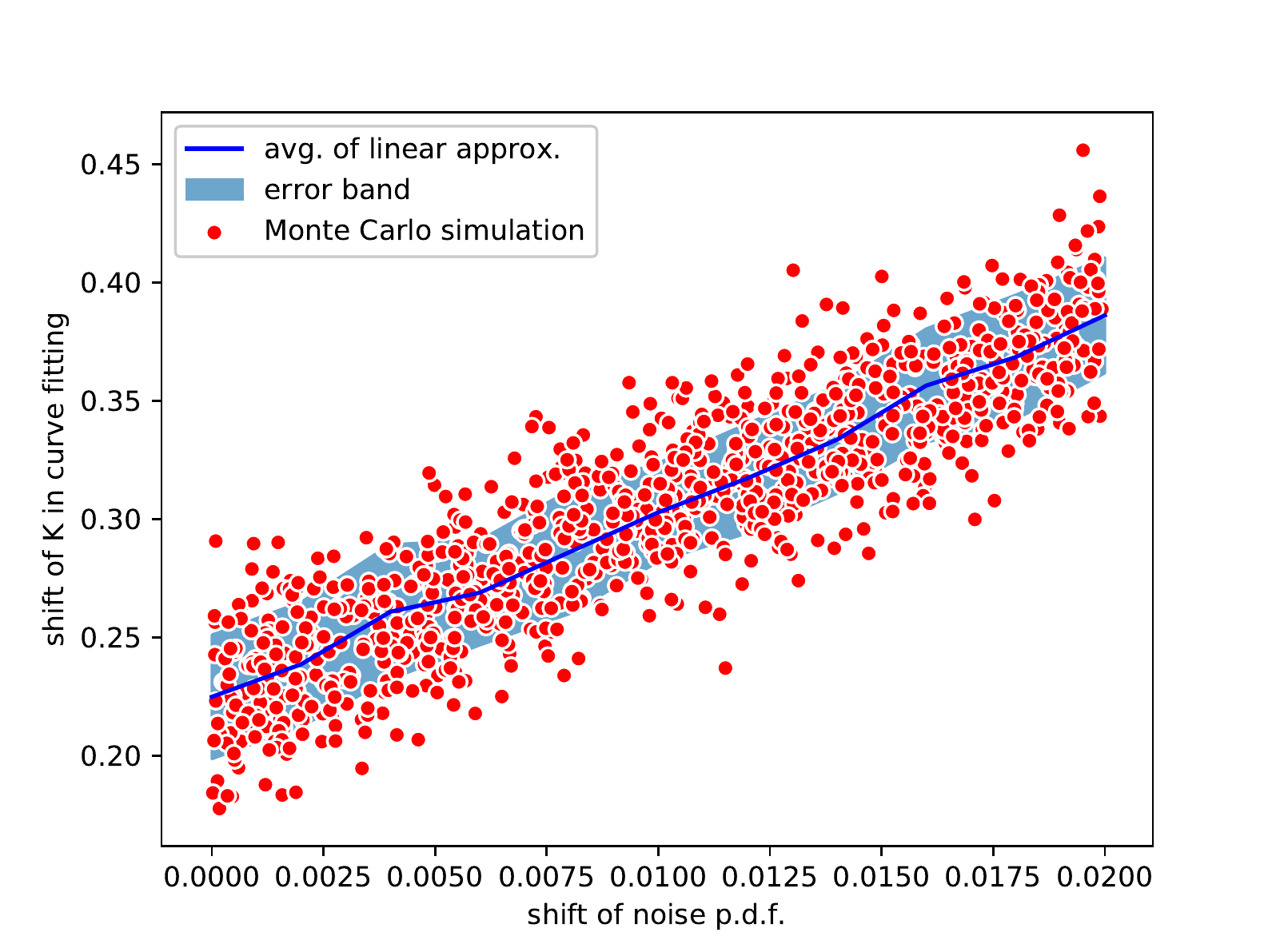}}		
	\subfigure[$t_0$ vs. shift of clipped Moyal]{
		\includegraphics[width=0.48\textwidth]{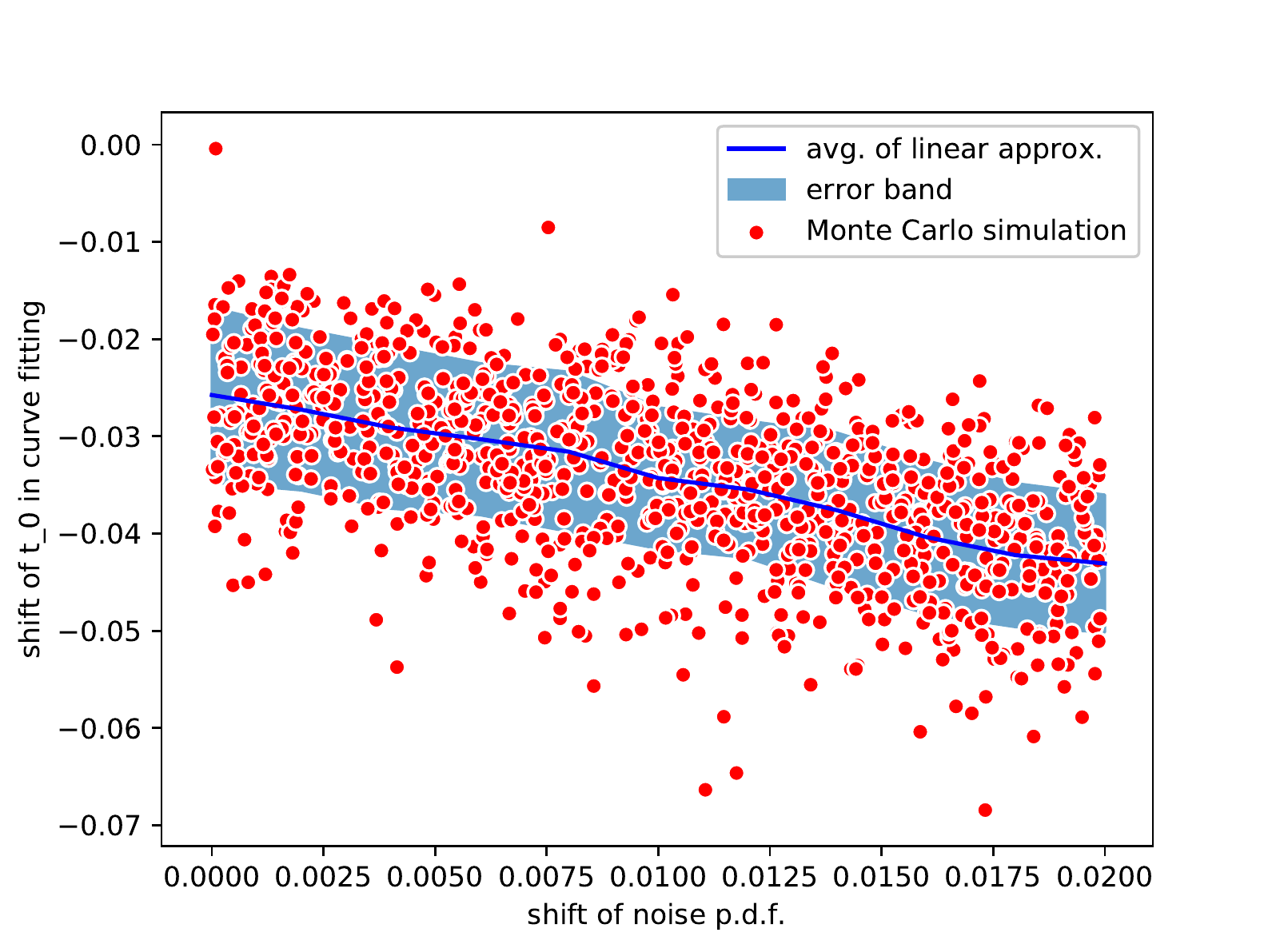}}
	\caption{\label{fig:curve-fitting-noise} A gathering of figures for noise simulations. Each figure compares the result from first-order (linear) approximations and the result from curve fitting directly.}
\end{figure}

\paragraph{Random noise} According to equation \ref{equ:dev-solve}, if the per-sample noise is Gaussian, the linear mapping will propagate the noise to the fitting parameters directly, so the distribution of fitting parameters will also be Gaussian. On the other hand, if the per-sample noise is not Gaussian, the linear mapping will work in a similar way. In order to study the distribution of fitting parameters for these non-Gaussian cases, we select two representative noise distributions, which are the crystal ball distribution \cite{gaiserappendix} and the Moyal distribution \cite{walck1996hand}. The former one has a long tail at the left-hand side and the latter one has a long tail at the right-hand side. Their probability density functions are:

\begin{equation}
	\text{crystal ball:  } f(x, \beta, m) =  \begin{cases}
	N \exp(-x^2 / 2),  &\text{for } x > -\beta\\
	N A (B - x)^{-m}  &\text{for } x \le -\beta
	\end{cases}
\end{equation}

\begin{equation}
	\text{Moyal:  } f(x) = \exp(-(x + \exp(-x))/2) / \sqrt{2\pi}
\end{equation}

\noindent For the crystal ball distribution, we choose $\beta = 2$, $m = 3$, shift its center to $[2.0, 4.0]$ and downscale it with 0.01. For the Moyal distribution, we shift its center to $[3.2, 6.4]$ and downscale it with 0.00625. In addition, in order to study the non-negative effect in detector electronics, we clip the noise to force the noise values to 0 if they become negative. The simulation results are shown in figure \ref{fig:curve-fitting-noise}. For each figure, we calculate the first-order approximations and run Monte Carlo simulation with a volume of 1000. It can be seen that although the noise distributions have strong non-Gaussian features, the distributions of fitting parameters have Gaussian shapes. The mean and standard deviation calculated from equation \ref{equ:dev-solve} can very well characterize the distributions from curve fitting. This implies that for medium size (33 in this case) of sampling points, distributions of fitting parameters show accordance to the law of large numbers, which is a statistical property of many independent random variables.

\paragraph{} In conclusion, the first-order approximations can describe curve fitting in a simple and convenient way. Different variations can be viewed as independent forces that drive the deviation of fitting parameters, and their relation is additive. This paves the way for the comparison in section \ref{sec:comparative-study} where we will demonstrate the potential of deep learning against this perspective.

\section{New approach --- deep learning method}
\label{sec:deep-learning}

Deep learning is a major breakthrough in recent years. It is based on neural networks, but its focus has shifted to building intricate network architectures for real-world applications (eg. image, voice, natural language, etc.). It started with image classification tasks \cite{krizhevsky2012imagenet, he2016deep} and spread to other domains \cite{hinton2012deep, shen2014learning} in artificial intelligence. Furthermore, it has been applied to high energy physics in recent literatures \cite{de2016jet, racah2016revealing, renner2017background, acciarri2017convolutional}. In the following subsections, we discuss how to use deep learning to solve the pulse timing problem.

\subsection{Deep learning basics}

The concept of neural networks is fundamental in deep learning. The basic element of a neural network is called a \emph{neuron}. A neuron has N inputs and one output. Besides, it has N weights and one bias as its parameters. It computes the products of the inputs and the weights in an element-wise manner, adds them together with the bias and uses a nonlinear activation function on the sum. Many similar neurons can act on the same inputs and form a layer. For a neural network with one intermediate layer, the output unit is also a neuron. The only intermediate layer is also called the hidden layer.

A deep neural network usually refers to a network with more than one hidden layer. By taking the output of the former layer as the input, hidden layers can be stacked. Increasing the depth of the network can gain additional power to extract structured features and reduce the number of parameters needed to approximate some functions. In general, neural networks have promising mathematical characteristics. They are supported by the universal approximation theorem \cite{hornik1989multilayer, cybenko1989approximation}, which states that neural networks can approximate mathematical functions with arbitrary precisions with enough neurons and layers.

One successful network structure is the \emph{convolutional neural network} \cite{krizhevsky2012imagenet}. It is based on the ideas of weights sharing and shift invariance. Instead of connecting a neuron to all inputs, we compute the output of a neuron in a vicinity (eg. a 2D patch) of the neuron. Besides, the weights to produce the output are shared across different places. By taking these measures, the parameters in a neural network can be greatly reduced and the efficiency can be improved dramatically.

To train a neural network, we need (input, label) pairs. The input is propagated through the neural network and compared to the label to compute a loss function. Then the loss is used to update the parameters of the whole network by the back propagation algorithm. The updating formula is usually based on the gradient descent method, i.e. descending the parameters in a direction which reduces the loss function. The loss function is usually the cross entropy along with the softmax function for a classification task \cite{krizhevsky2012imagenet}, and the mean square error and its derivatives for a regression task.

\subsection{The potential of deep learning --- a comparative study}
\label{sec:comparative-study}

With the knowledge above, it is ready to discuss the potential of deep learning and compare it to the curve fitting method. The study is carried out in the aspect of variations in section \ref{sec:quan-analysis}.

\paragraph{Long-term drift} From the analysis in section \ref{sec:curve-fitting}, the long-term drift will introduce a bias to the fitting parameters. In a large detector system, correcting the bias is a tremendous task, and even impractical in some cases. For one thing, unlike the discussion in section \ref{sec:curve-fitting-sim}, system parameters are hidden in the function and sometimes have very sophisticated forms. For another, the non-uniformity of different cells makes the problem even more complicated. Furthermore, if we view the bias in the non-Gaussian noise as a kind of long-term drift, the total effect is a mixture of several aspects. To tackle the bias challenge, we can use a regression neural network to fix the influence of the long-term drift to the fitting parameters. Without loss of generality, we can assume that the last layer of the neural network has the form $y = f(\bm{x}; \bm{w}, b) = \sum_i w_i \cdot x_i + b$. Since the last layer has a bias parameter $b$, if there is a persistent shift in the system, this shift will be counteracted by the bias parameter $b$ through the training process. As long as the training label is sufficiently accurate, the bias can be greatly reduced by the neural network.

\paragraph{Short-term change} For curve fitting, the short-term change has a direct impact on the precision of the fitting parameters. In equation \ref{equ:dev-solve}, it can be seen that the event-by-event variations of the system parameters $\bm{\theta}$ will result in the fluctuation of the fitting parameters. The primary cause for this phenomenon is that curve fitting treats each set of ADC samples as an independent and complete set of features. However, different sets of ADC samples belong to the same function family, and an overall understanding of the function family is beneficial to the explanation of the individual set of features. The optimization of neural networks is such a global process which is helpful to establish the overall understanding. To see this point, we can rewrite the mapping of the neural network as:

\begin{equation}
	\bm{\beta^\prime} = \bm{g}(\bm{f}(\bm{t}; \bm{\beta}, \bm{\theta}) + \bm{n}; \bm{W}, \bm{B})
\end{equation}

\noindent where $\bm{f}(\bm{t}; \bm{\beta}, \bm{\theta}) = (f(t_1; \bm{\beta}, \bm{\theta}), f(t_2; \bm{\beta}, \bm{\theta}), \ldots, f(t_n; \bm{\beta}, \bm{\theta}))$ is the vector of sampling points, and $\bm{W}, \bm{B}$ are the weights and biases of the neural network. When we optimize the model, the training label will change consistently with the underlying fitting parameters $\bm{\beta}$ but remain the same when system parameters $\bm{\theta}$ vary. As a result of training, the weights $\bm{W}$ and biases $\bm{B}$ of the neural network follow a gradient descent direction so that the change of $\bm{\beta^\prime}$ is proportional to the change of $\bm{\beta}$ but orthogonal to the change of $\bm{\theta}$. In other words, training increases the sensitivity to variations of fitting parameters and reduces the sensitivity to variations of system parameters. In this way, the influence of the short-term change can be greatly alleviated.

\paragraph{Random noise} We have already analyzed the Gaussian noise with the accurate fitting function in section \ref{sec:adv-curve-fitting}. Here we focus on the noise with more complex forms. According to the central limit theorem, the distributions of fitting parameters will take Gaussian shapes when noise is presented. This is a degenerative process and could loss original information. To help understand the claim, we might think of the development of modern physics. When the instrumentation was not so advanced, people could only observe macro phenomenons, which were normally distributed according to statistical laws. Once the hardware condition had improved, people could measure the micro mechanisms, and the fine structures could be found. In our problem, curve fitting does not utilize the information in each time point sufficiently, and the loss of information can not be retrieved. On the other hand, we already know that neural networks have micro structures. This offers an opportunity to achieve better performance than curve fitting in the non-Gaussian settings. Since the nonlinear mapping in the activation function can implement a complicated function family, it is possible to use neural networks to retrieve the origin information from noisy inputs.

\paragraph{} In conclusion, deep learning is a good alternative to the traditional curve fitting method in terms of drift, change and noise when used in an appropriate way.

\subsection{Network architecture}
\label{sec:nn-arch}

In this part, we will discuss the implementation issues of deep learning in the specific pulse timing problem. Although neural networks are promising according to the analysis in section \ref{sec:comparative-study}, it does not mean that any structure will perform well. When facing a new problem, practitioners need to customize the network structure to make it suitable for the problem settings.

\begin{figure}[htbp]
	\centering	
	\includegraphics[width=0.9\textwidth]{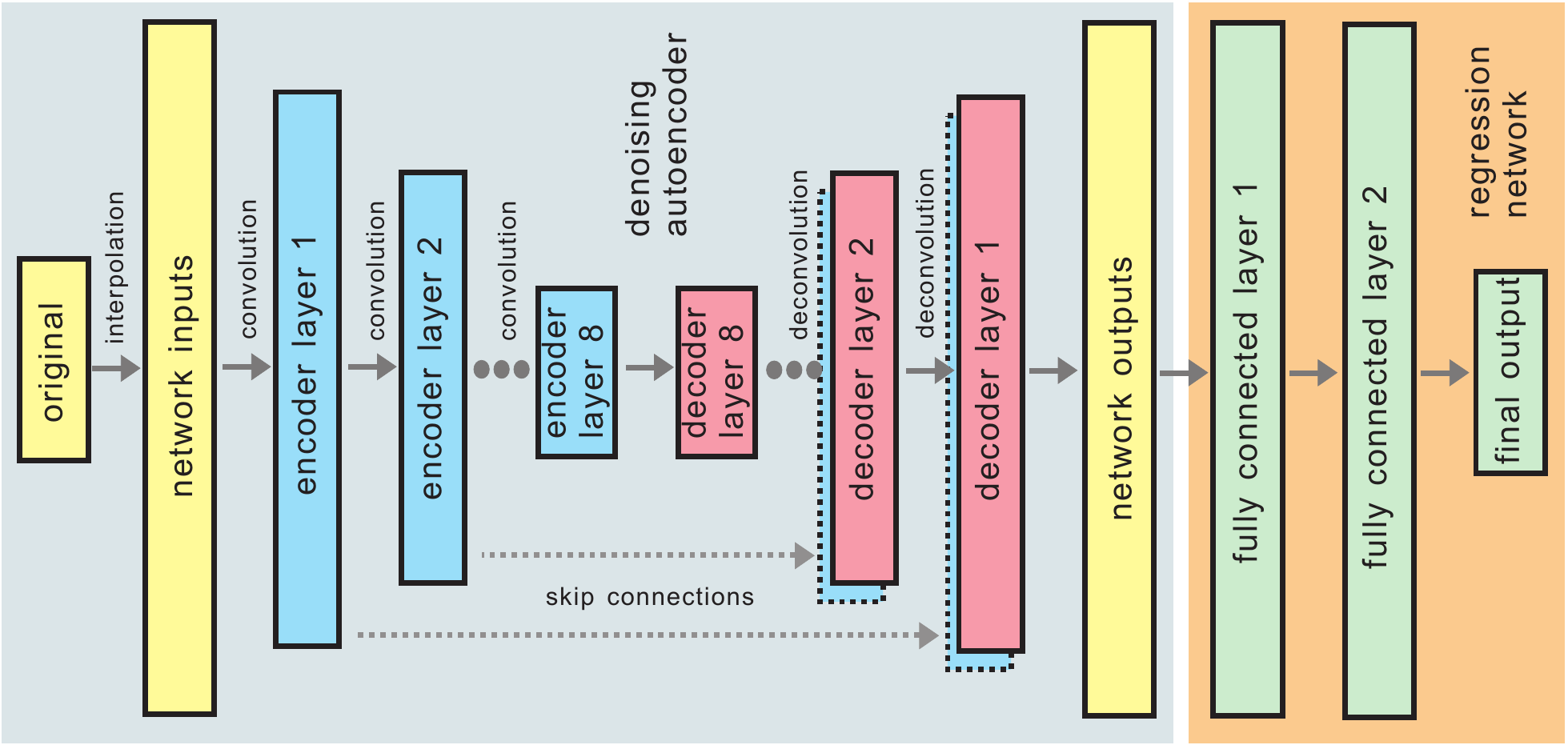}		
	\caption{\label{fig:network} A diagram of the network architecture.}
\end{figure}

We design our network architecture based on the ideas from \cite{isola2017image, kuleshov2017audio}. A diagram of the adopted architecture is shown in figure \ref{fig:network}. In principle, the network is comprised of two parts, a denoising autoencoder and a regression network.

The denoising autoencoder \cite{vincent2010stacked} is a network which tries to recover the original unstained input from its noisy version. A typical autoencoder is made up of a pyramid structure which performs feature extraction (encoding), and an inverted pyramid structure which restores original data (decoding). We add following features to the prototype of the autoencoder to improve its performance:

\begin{enumerate}
	\item \emph{Convolution and deconvolution}. In the encoder layers and decoder layers, we use convolution \cite{krizhevsky2012imagenet} and deconvolution \cite{noh2015learning} operations to replace the fully-connected layers. These operations can utilize the locality of input features and extract structured patterns from data. In the convolution, we use many groups of parameters (called a \emph{filter} or \emph{kernel}) to compute the output (called a \emph{feature map}). Each filter has its own weights and bias, and it moves across the input feature map to produce an one-dimensional output. Many filters will result in a feature map with many channels of one-dimensional data. In the deconvolution, the operation between the input and the output is transposed. For the same stride and padding, the output shape of deconvolution operations will be the same as the input shape of the corresponding convolution operations.
	
	\item \emph{Skip connections}. Optimizing a deep neural network suffers from problems of vanishing/exploding gradients. Even when these problems are handled by normalization, a degrading problem still affects the performance of the model. In \cite{he2016deep}, a dedicated structure called the \emph{residual network} was suggested to solve the problem. In view of this work, we implement skip connections between the encoder layers and decoder layers to overcome the issues when training a deep network. Except the last layer, every layer in the encoder is directly copied to the corresponding layer in the decoder. At the decoding side, the channels from the encoder and the channels from the main passage of the network are concatenated. In this way, the relation between long-range layers is preserved so that it is easier for the network to learn valuable features from the input.
	
	\item \emph{Leaky ReLU}. The Rectified Linear Unit (ReLU) \cite{nair2010rectified} is a kind of activation function which is widely used in deep learning. Since ReLUs force the output to become zero when the input is negative, it blocks the flow of information for a considerable amount of neurons in a network. The leaky ReLU \cite{xu2015empirical} is proposed to solve the problem. Unlike ReLUs, the leaky ReLU has a gradual slope at the negative x-axis. It has a non-zero gradient even when the input is negative. In our network, we use leaky ReLUs in the encoder layers.
\end{enumerate}

\begin{table}[htbp]
	\centering
	\caption{Specification for the denoising autoencoder.}
	\label{tab:denosing-autoencoder}
	\scriptsize
	\begin{tabular}{|lllll|}
		\hline
		\multicolumn{5}{|c|}{Convolution} \\
		No. & stride & filter width & out channels & leaky ReLU \\
		\hline
		1 & 2 & 4 & 64 & No ReLU \\
		2 & 2 & 4 & 128 & Yes (0.2) \\
		3 & 2 & 4 & 256 & Yes (0.2) \\
		4 & 2 & 4 & 512 & Yes (0.2) \\
		5 & 2 & 4 & 512 & Yes (0.2) \\
		6 & 2 & 4 & 512 & Yes (0.2) \\
		7 & 2 & 4 & 512 & Yes (0.2) \\
		8 & 2 & 4 & 512 & Yes (0.2) \\
		\hline
	\end{tabular}
	{\begin{tabular}{|lllll|}
			\hline
			\multicolumn{5}{|c|}{Deconvolution} \\
			No. & stride & filter width & out channels & dropout \\
			\hline
			8 & 2 & 4 & 1024 & Yes (0.5) \\
			7 & 2 & 4 & 1024 & Yes (0.5) \\
			6 & 2 & 4 & 1024 & Yes (0.5) \\
			5 & 2 & 4 & 1024 & No \\
			4 & 2 & 4 & 512 & No \\
			3 & 2 & 4 & 256 & No \\
			2 & 2 & 4 & 128 & No \\
			1 & 2 & 4 & 1 & No \\
			\hline
	\end{tabular}}
\end{table}

Specifically, the denoising autoencoder is a network with 8 $\times$ 2 layers. First, we use cubic (or quadratic) interpolation to stretch the original input to the desired length. Then it goes through the network to get the output. The specification in detail is shown in table \ref{tab:denosing-autoencoder}. In the convolution part, leaky ReLUs are used except the first layer. The number is the parentheses represents the slope at the negative x-axis. In the deconvolution part, the output channels include channels from skip connections. Dropout \cite{srivastava2014dropout} is a regularization method to prevent overfitting. We use dropout in the first three layers. The number in the parentheses represents the dropout ratio.

Upon the denoising autoencoder, we add a regression network to directly output the parameters of interest. The structure of the regression network is a traditional feedforward network with 2 hidden layers. Each layer, with 512 neurons, is fully connected to its input. A \emph{softmax} layer is used between the denoising autoencoder and the regression network.

Training such a network can be divided into the following two steps:

\begin{enumerate}
	\item \emph{Autoencoder pre-training}. It is strongly recommended to pre-train the denoising autoencoder as the first step of the training process. Based on the function of the autoencoder, we need to estimate the form of noise and generate (noisy input, unstained input) pairs as the (input, label) to train the network. To be more specific, first we randomly generate a set of sampling points according to the pulse function. Then we add per-sample noise to the sampling points according to the probability distribution of the estimated form of noise. If the expression of the short-term change is known, it can also be used. Actually, only a rough estimate can improve the final performance significantly (see section \ref{sec:exp-results}). In this stage, only simulation data is used.
	
	\item \emph{End-to-end finetuning}. After pre-training, we can use experimental data (if available) to make an end-to-end finetuning of the whole network. A precise label indicating the ground-truth parameter is used at the far-end of the network to generate a loss function. There are two options in finetuning. The first option is to keep the autoencoder unchanged and only finetune the regression network. If there are no distinct changes in the pulse function compared to the pre-training stage, this option can be used. The second option is to finetune the whole network together. For this option, the pre-trained network only works as an optimal start point for finetuning, and the capacity of the model is larger (which also implies overfitting issues).
\end{enumerate}

\subsection{Simulation studies}
\label{sec:nn-sim}

In this part, we run simulations of the proposed neural network regarding the variations discussed in section \ref{sec:quan-analysis}. Since the advantage of the neural network model on the long-term drift is evident according to the discussion in section \ref{sec:comparative-study}, we do not run simulations for this kind of variation.

In order to study the variations, first we need to generate the simulation dataset. The pulse function is the same as section \ref{sec:curve-fitting-sim}. In the following simulations, we choose $K$ uniformly sampled in the range $[2.56, 5.12]$ and $t_0$ uniformly sampled in the range $[-0.9, 0.1]$. The reference values for $\tau_p$ and $b$ are 2.0, 0.1 respectively. The pulse for the noisy input (or the input with short-term change) is sampled from $t = 0.0$ to $t = 3.2$ with a period of $0.1$. We drop the last point when training, so there are 32 points. The same pulse for the label is sampled at a super-resolution ratio of 8 in the same interval, so there are a total of 256 points. We gather the simulation samples into two separate datasets. The training dataset has 40000 samples and the test dataset has 10000 samples.

To calculate the timing resolution, we test different methods on the test dataset and get the predicted values of the start time $t_0$. For curve fitting, the predicted values are the fitting parameters. For regression networks, the predicted values are the outputs of the networks. Then we use the difference between the predicted values and the ground-truth values to make a Gaussian fit. The standard deviation of the Gaussian fit is a measure of the timing resolution.

\begin{figure}[htbp]
	\centering
	\subfigure[A typical figure of the inputs, the outputs and the targets (label) of the autoencoder.]{			
		\includegraphics[width=0.48\textwidth]{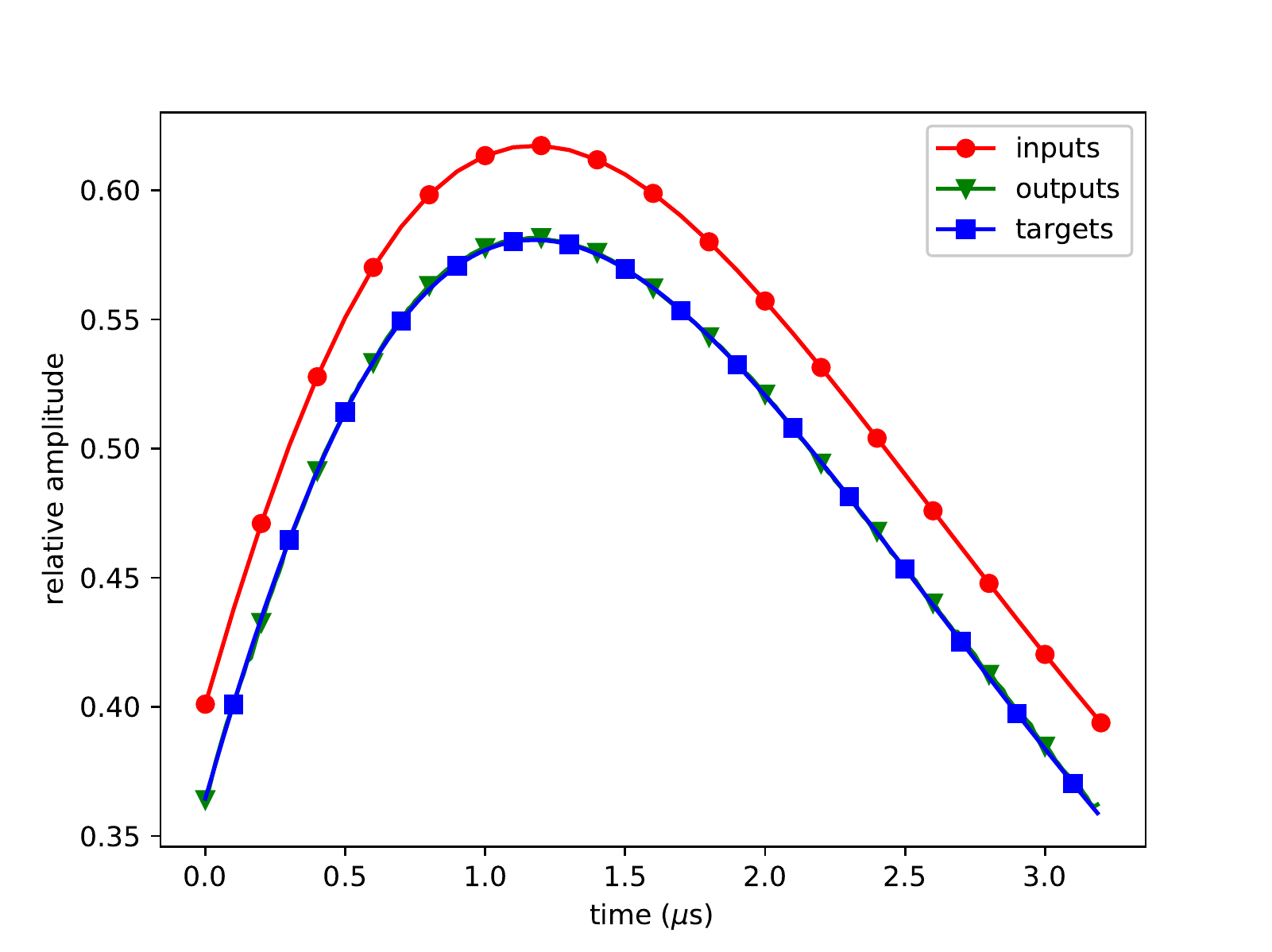}}
	\hfill
	\subfigure[The RMS of amplitude between the inputs/outputs and the ground-truth targets. The figure is plotted on the statistics of the whole test dataset.]{
		\includegraphics[width=0.48\textwidth]{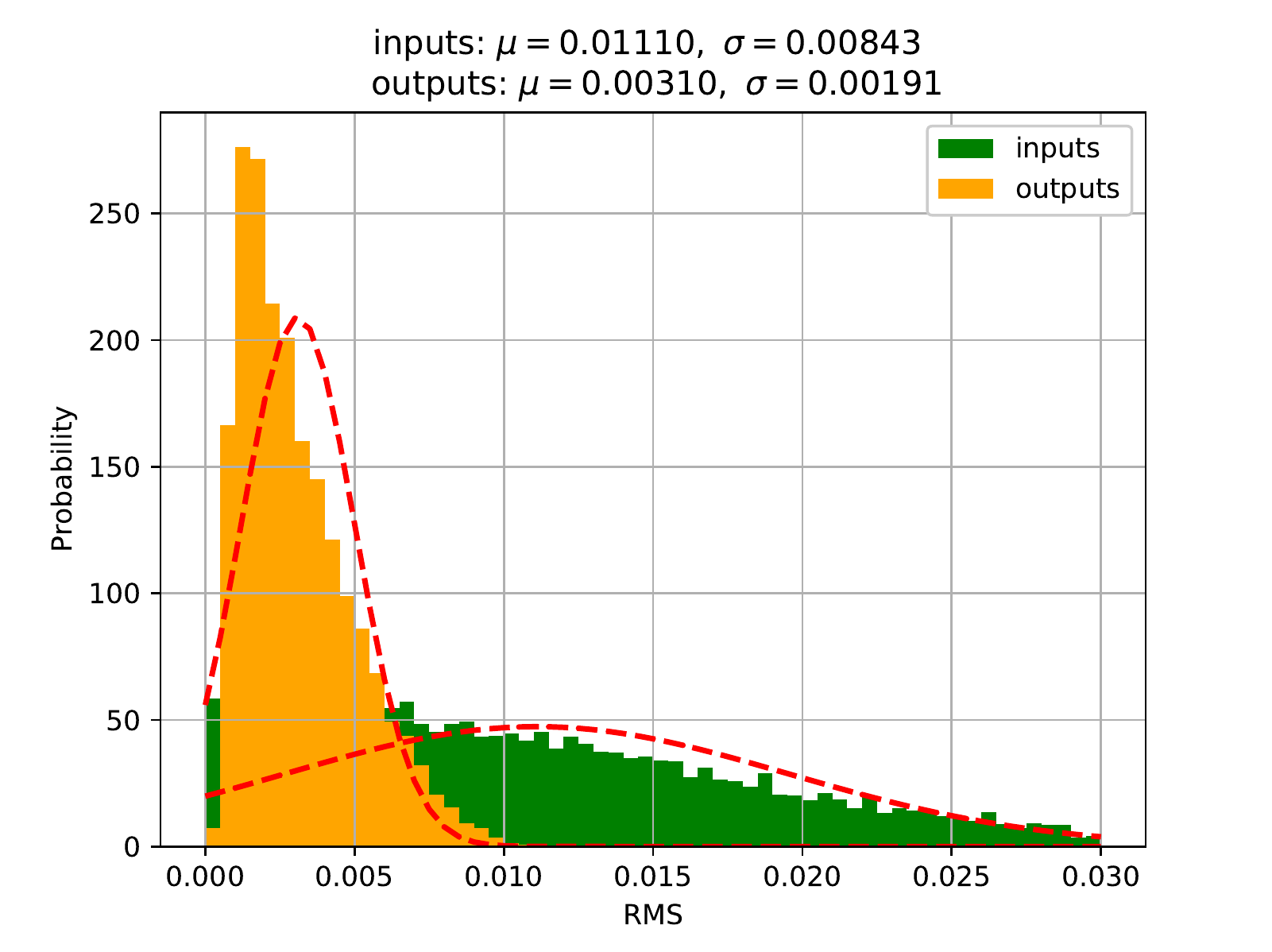}}
	\caption{\label{fig:nn-short-term-change} The simulation results of the denoising autoencoder for the short-term change. (\emph{left}) We choose a sample in the test dataset and plot the noisy input, the denoising outcome and the training label. (\emph{right}) We calculate the Root Mean Square (RMS) between the inputs/outputs of the neural network and the ground-truth label for each sample in the test dataset. Then we make a Gaussian fit for all the samples and plot the figure.}
\end{figure}

\begin{table}[htbp]
	\centering
	\caption{Simulation results for the short-term change. The table compares different neural network models with curve fitting.}
	\label{tab:nn-short-term-change}
	\small
	\begin{tabular}{|cccc|}
		\hline model & note & converged & timing resolution ($\mu$s) \\
		\hline fitting original data & --- & --- & 0.01217 \\
		fitting autoencoder outputs & only base network & --- & 0.00296 \\
		regression net v1 & base network fixed & successful & 0.00303 \\
		regression net v2 & base network trainable & successful & 0.00182 \\
		\hline
	\end{tabular}
\end{table}

\paragraph{Short-term change} To study the effects of the short-term change, we introduce the baseline shift, i.e. the variations of the pedestal $b$. The baseline shift is a common type of the short-term change especially when the event rate is high so that nearby events will interplay. To construct the dataset, first we add the same shift to all sampling points in an event. The shift is randomly sampled from a Gaussian distribution with 0.1 mean and 0.014 standard deviation. The training targets of the denoising autoencoder are set to have the pedestal $b = 0.1$, which is the standard value used in curve fitting. The results are shown in figure \ref{fig:nn-short-term-change} and table \ref{tab:nn-short-term-change}. In the left figure, we can see that although the pedestal $b$ and the amplitude $K$ are both random and have high correlation, the denoising autoencoder can effectively perceive the change in the pedestal and cancel the change. The right figure shows the distribution of RMS based on the statistics of the whole test dataset. The average of RMS is reduced from 0.01110 to 0.00310 by a factor of 3.58. In the table, we compare the timing resolution achieved by curve fitting and neural networks. In the first two lines, it can be seen that fitting the outputs of the denoising autoencoder is better than fitting original data, which demonstrates the effectiveness of the neural network structure. The result of the regression network v1 when the base network is fixed is slightly worse than fitting the outputs of the denoising autoencoder. The best result (1.82 ns) comes with the regression network v2 when the base network is trainable. It outperforms curve fitting results significantly. It implies that, for the short-term change, when we choose a proper start point and finetune the whole network, the result can be even better than the autoencoder alone.

\paragraph{Random noise} We analyze two representative kinds of noise: the Gaussian noise and the clipped Moyal noise (see section \ref{sec:curve-fitting-sim}).

\begin{figure}[htbp]
	\centering
	\subfigure[A typical figure of the inputs, the outputs and the targets (label) of the autoencoder.]{			
		\includegraphics[width=0.48\textwidth]{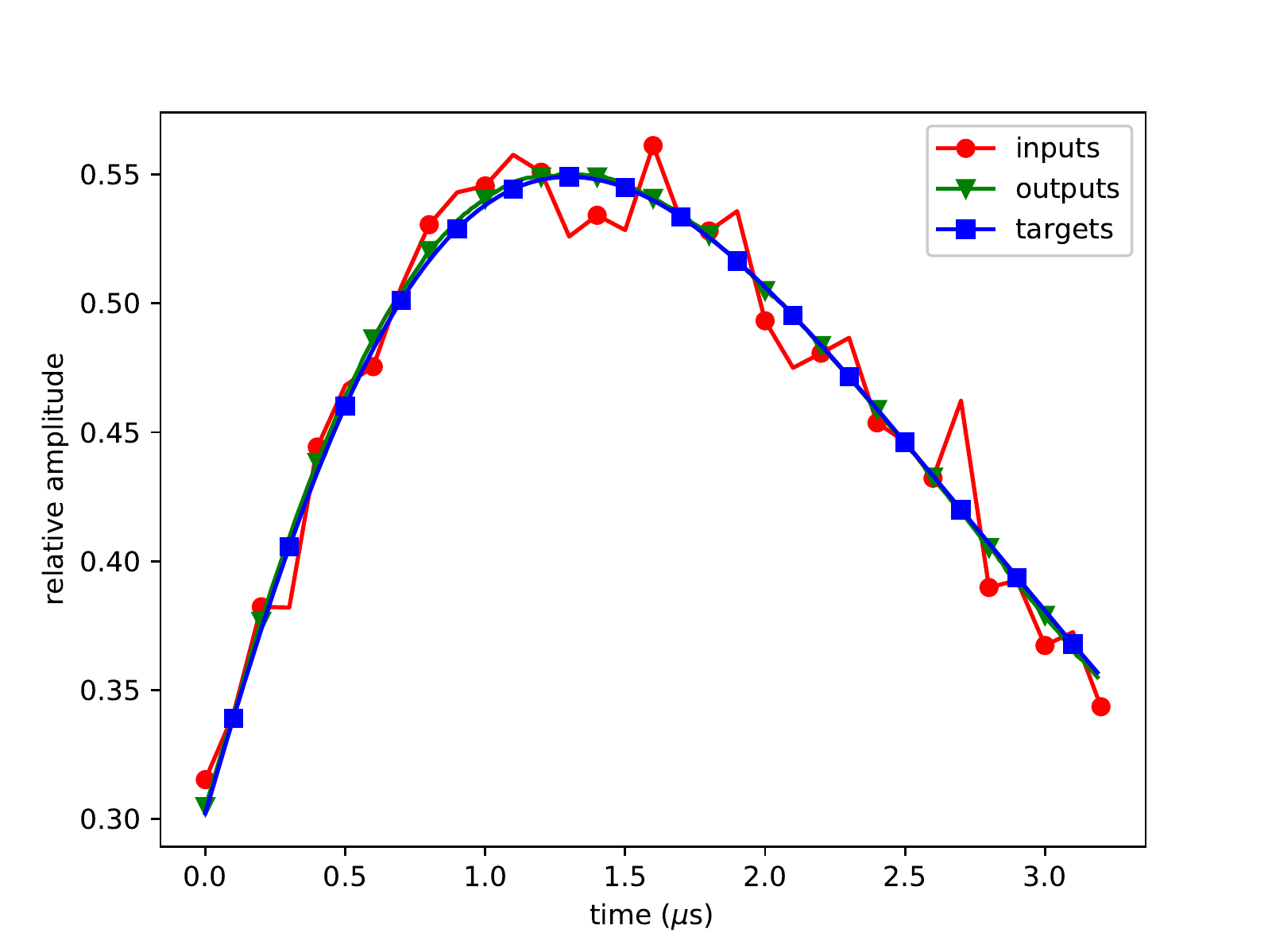}}
	\hfill
	\subfigure[The RMS of amplitude between the inputs/outputs and the ground-truth targets. The figure is plotted on the statistics of the whole test dataset.]{
		\includegraphics[width=0.48\textwidth]{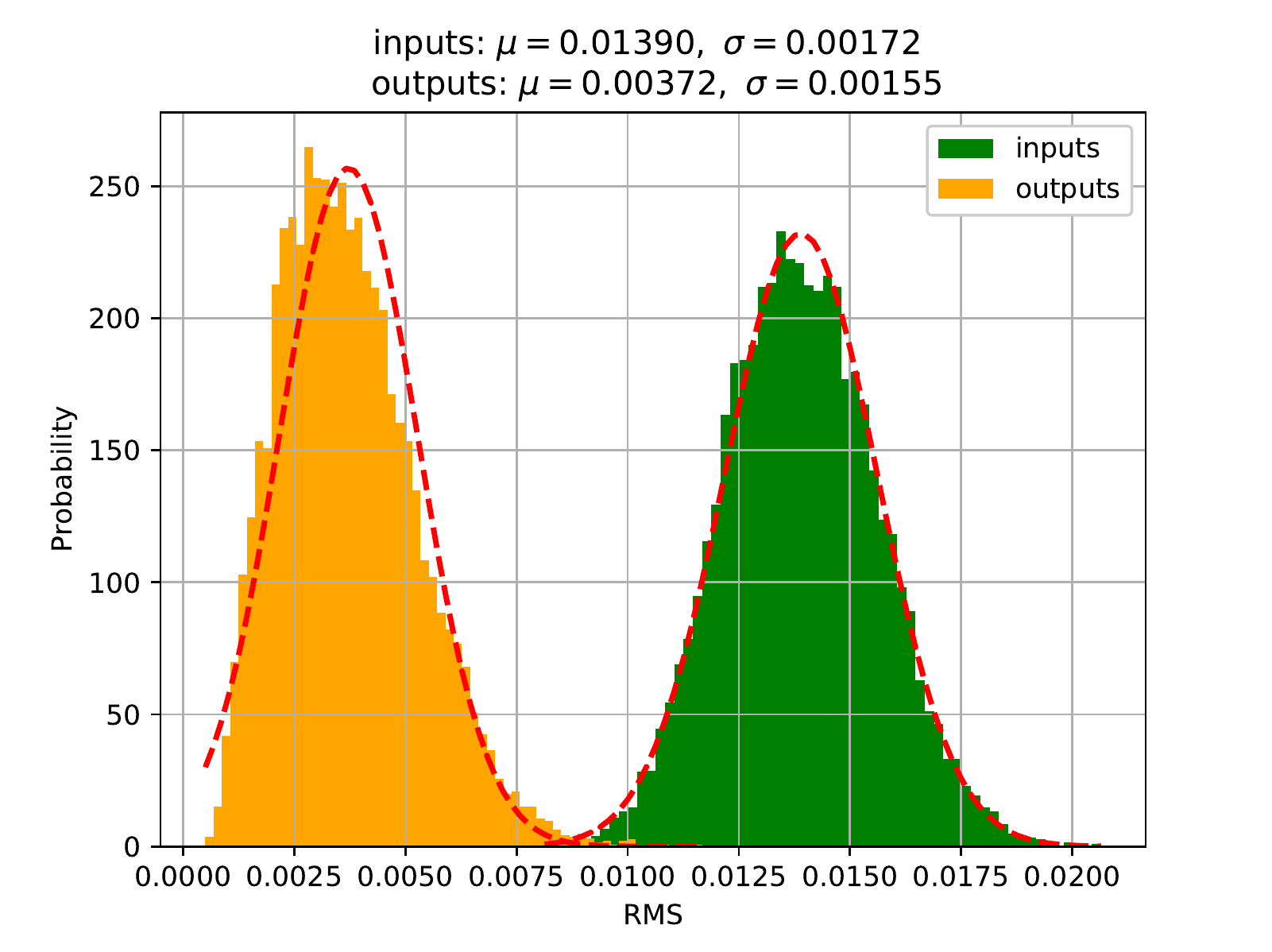}}
	\caption{\label{fig:nn-gaussian-noise} The simulation results of the denoising autoencoder for the Gaussian noise. The figures are plotted in the same way as figure \ref{fig:nn-short-term-change}.}
\end{figure}

\begin{table}[htbp]
	\centering
	\caption{Simulation results for the Gaussian noise. The table compares different neural network models with curve fitting.}
	\label{tab:nn-gaussian-noise}
	\small
	\begin{tabular}{|cccc|}
		\hline model & note & converged & timing resolution ($\mu$s) \\
		\hline fitting original data & maximum likelihood estimator & --- & 0.01206 \\
		only regression net & no base network & failed & 0.26756 \\
		fitting autoencoder outputs & only base network & --- & 0.01249 \\
		regression net v1 & base network fixed & successful & 0.01530 \\
		regression net v2 & base network trainable & successful & 0.01261 \\
		\hline
	\end{tabular}
\end{table}

In the first place, we add the Gaussian noise with zero mean and 0.014 standard deviation. This introduces a noise ratio $\frac{\text{noise std. deviation}}{\text{average amplitude}} \approx 2.7\%$. In order to give results with visual impacts, the noise ratio we choose is significantly larger than reality. The results are shown in figure \ref{fig:nn-gaussian-noise} and table \ref{tab:nn-gaussian-noise}. In the left figure, we can see that although the input points have obvious noise, the noise is suppressed by the denoising autoencoder so that the output points are approximating the label. In this sample and the majority of samples in the test dataset, the difference between the output points and the label is very slight. The right figure displays the statistics of the whole test dataset. The average of the noise RMS is reduced from 0.01390 to 0.00372 by a factor of 3.74. In the table, we use three neural network models and compare their performance with curve fitting. Since the Gaussian noise is the most common case, in the analysis we add the regression network alone for comparison. According to section \ref{sec:adv-curve-fitting}, fitting original data gives the result of the maximum likelihood estimator which is the theoretical lower bound. It can be seen that the network architecture is important to achieve the optimal performance. When we use only the regression network, the model fails to converge and gives a result worse than the sampling period. However, when we use the autoencoder-regression network architecture, the model can converge successfully. The best result of neural networks comes from the regression network v2 with the base network trainable. This shows the advantage of the model capacity in the problem.

\begin{figure}[htbp]
	\centering
	\subfigure[A typical figure of the inputs, the outputs and the targets (label) of the autoencoder.]{			
		\includegraphics[width=0.48\textwidth]{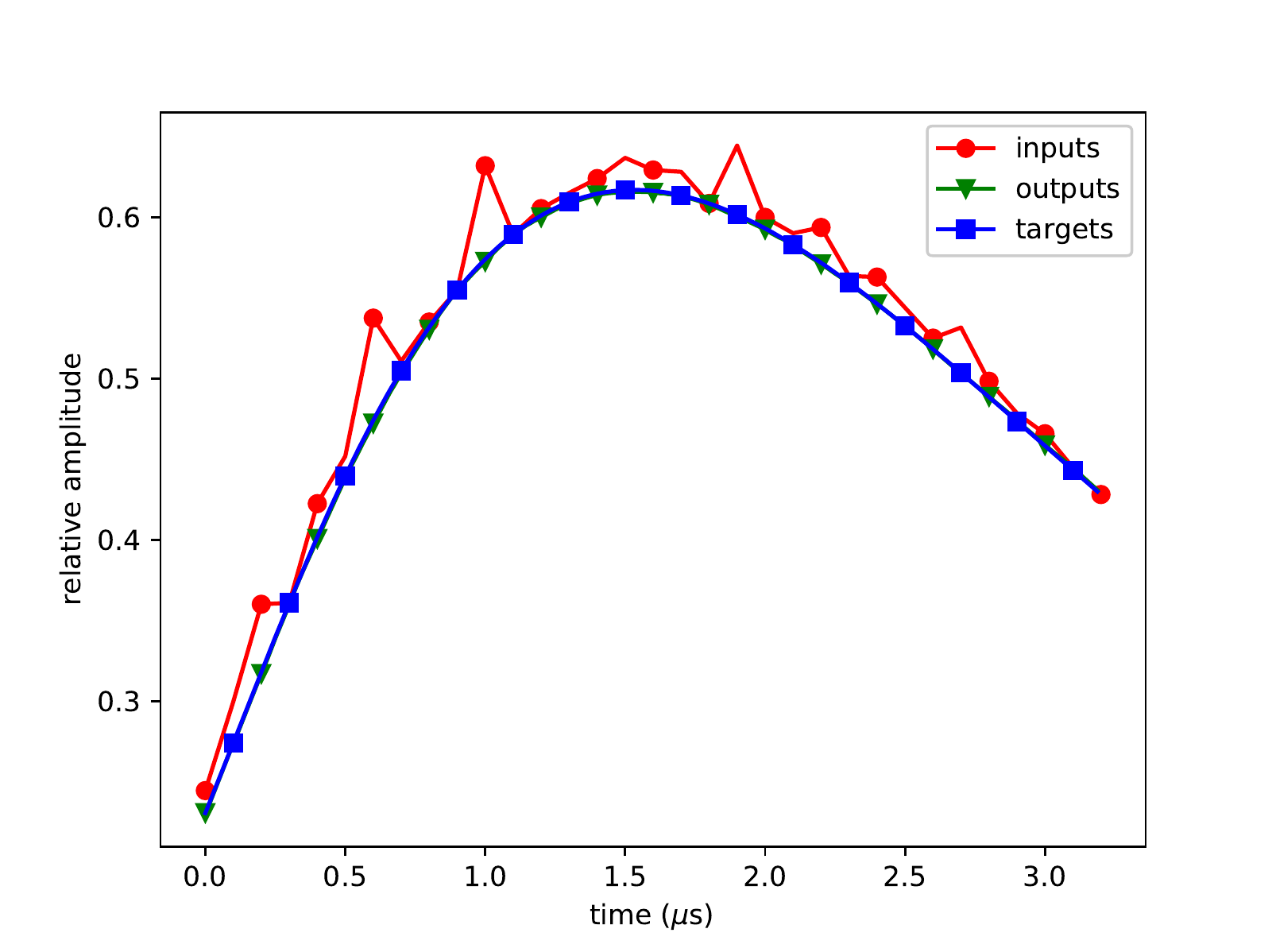}}
	\hfill
	\subfigure[The RMS of amplitude between the inputs/outputs and the ground-truth targets. The figure is plotted on the statistics of the whole test dataset.]{
		\includegraphics[width=0.48\textwidth]{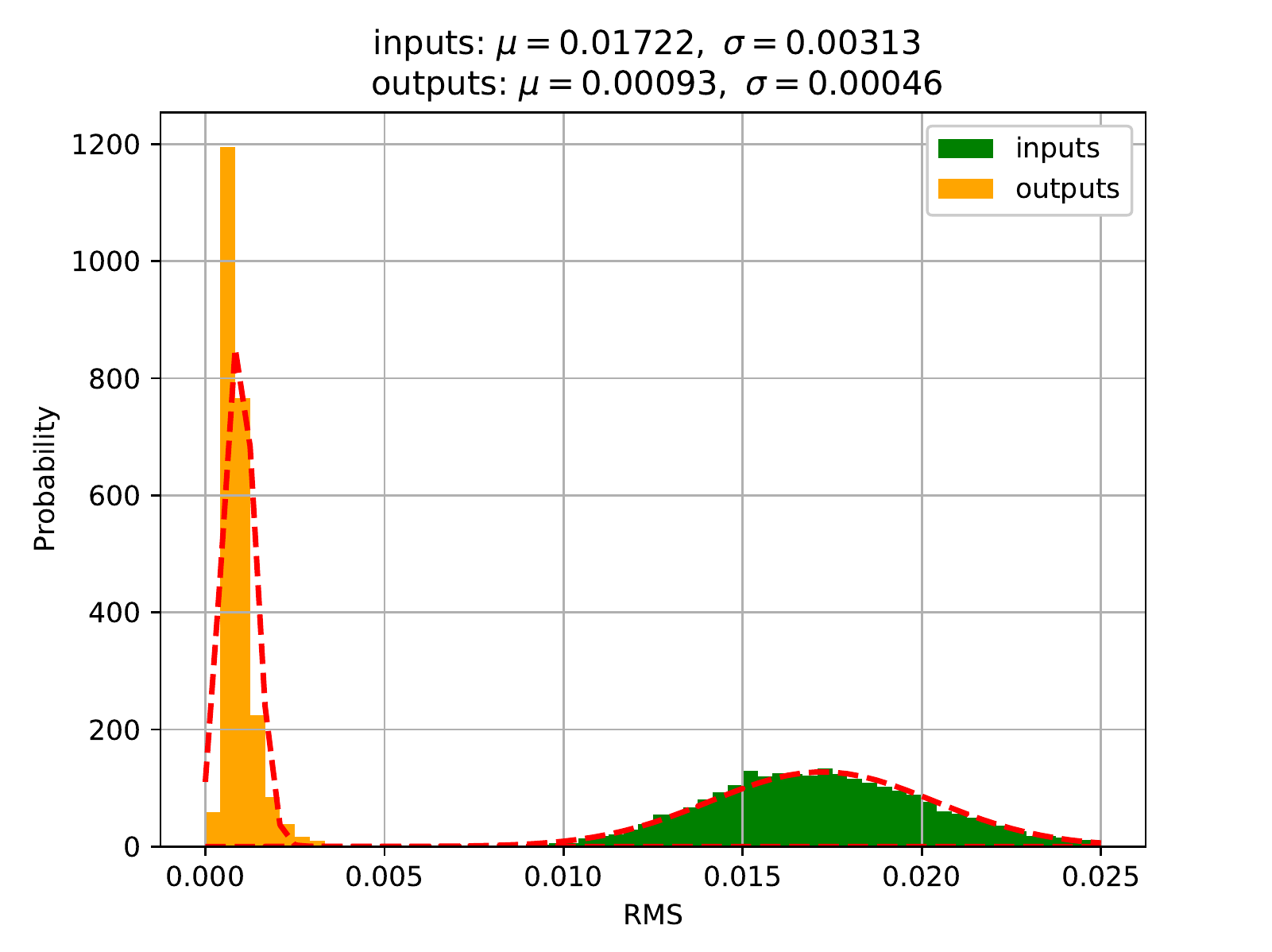}}
	\caption{\label{fig:nn-moyal-noise} The simulation results of the denoising autoencoder for the clipped Moyal noise. The figures are plotted in the same way as figure \ref{fig:nn-short-term-change}.}
\end{figure}

\begin{table}[htbp]
	\centering
	\caption{Simulation results for the clipped Moyal noise. The table compares different neural network models with curve fitting.}
	\label{tab:nn-moyal-noise}
	\small
	\begin{tabular}{|cccc|}
		\hline model & note & converged & timing resolution ($\mu$s) \\
		\hline fitting original data & --- & --- & 0.01203 \\
		fitting autoencoder outputs & only base network & --- & 0.00324 \\
		regression net v1 & base network fixed & successful & 0.00463 \\
		regression net v2 & base network trainable & successful & 0.00487 \\
		\hline
	\end{tabular}
\end{table}

In the second place, we analyze the clipped Moyal noise. The original Moyal distribution is shifted to location 0.004, rescaled with 0.006 and then clipped for noise generation. Again, the noise is more intense than reality. The results are shown in figure \ref{fig:nn-moyal-noise} and table \ref{tab:nn-moyal-noise}. In the left figure, we can see that the unique structure of the denoising autoencoder can very well get the clue of the ground-truth target from the noisy input. To further illustrate the idea, we plot the distribution of RMS on the test dataset in the right figure. The average of the noise RMS is reduced from 0.01722 to 0.00093 by a factor of 18.52. This exceeds the results from former simulations. In the table, we compare the timing resolution between curve fitting and neural networks. In the first two lines, it can be seen that curve fitting with the denoising autoencoder alone can improve the timing resolution significantly. Besides, when regression networks are added, the models can successfully converge and show competitive results. In this case, keeping the base network fixed (regression network v1) is slightly better than making the base network trainable (regression network v2), which demonstrates the good baseline provided by the autoencoder.

\paragraph{} To conclude the simulation results, the network architecture proposed in section \ref{sec:nn-arch} can very well tackle the non-ideal conditions. Finetuning the whole network together can achieve results better than fitting the outputs of autoencoder when the short-term change is applied, but slightly worse when the random noise is applied. Finetuning the regression network alone can sometimes achieve better results than finetuning the whole network, especially when the base network is accurate. In experimental conditions, it is not always possible to provide exact training targets for the denoising autoencoder as in the simulations. Thus, finetuning the regression network with the precise time label is vital to improve the performance of the whole network.

\section{Experimental results}
\label{sec:exp-results}

\begin{figure}[htbp]
	\centering
	\includegraphics[width=0.9\textwidth]{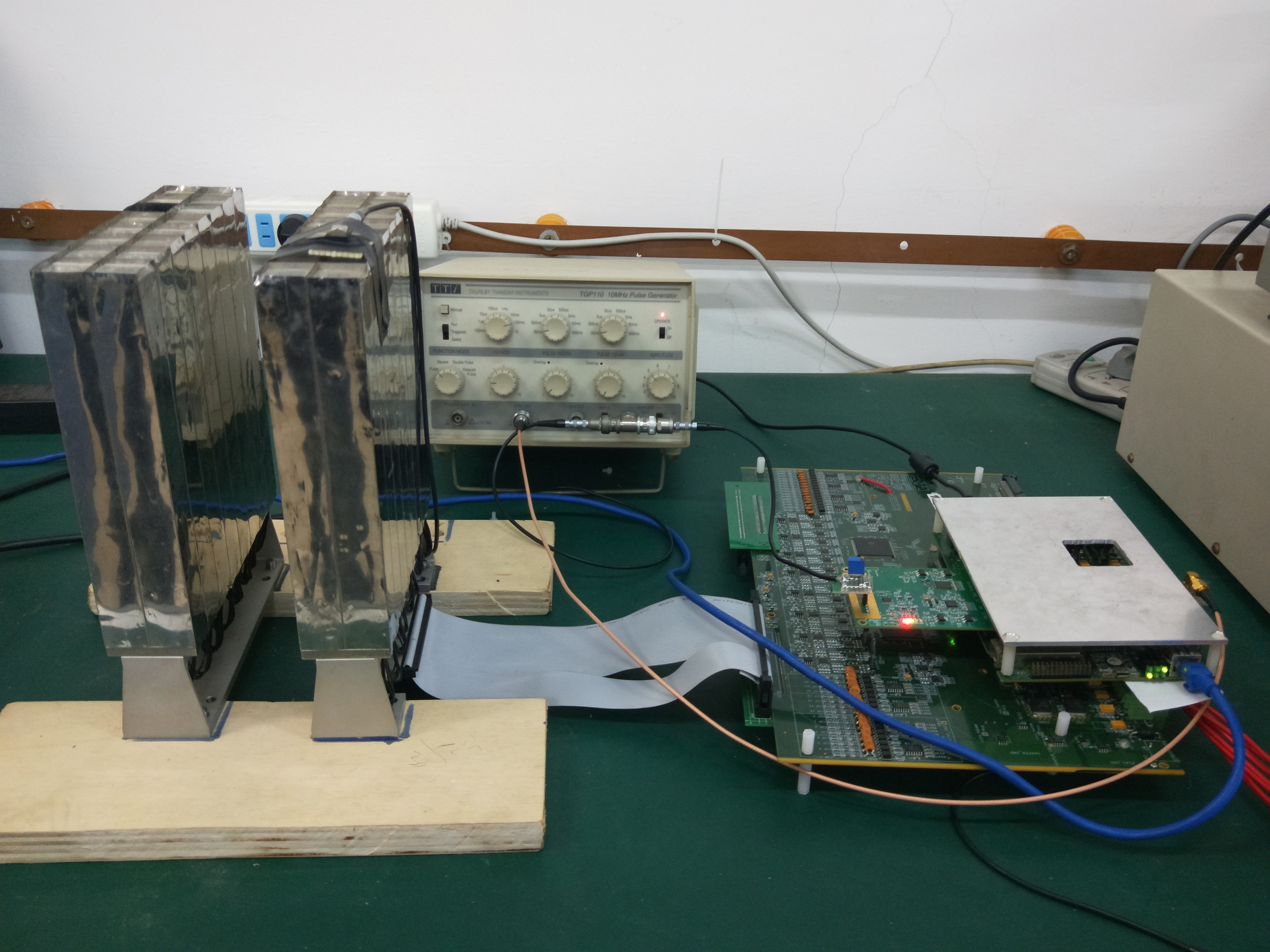}
	\caption{\label{fig:hardware-platform} A photograph of the hardware test platform with the PHOS detector, AD9656 data acquisition board and HPDAQ.}
\end{figure}

We build a hardware test platform to study the pulse timing in the real-world environment. A photograph of the platform is shown in figure \ref{fig:hardware-platform}. The test platform is based on the PHOS detector (section \ref{sec:alice-phos}). We use a pulse generator to produce pulses with the $\sim$50 ns width and the $\sim$10 Hz frequency. This pulse signal drives a LED to produce light for the PHOS crystal. The scintillation is collected by the APD and passed to the CSA. Then it is transmitted to the CR-RC2 shaper on the FEE card. The output of the CR-RC2 shaper is hardwired to the AD9656 data acquisition board, which is connected to the HPDAQ motherboard for TCP/IP communications. The AD9656 is a 4-channel ADC chip with 2.8 V dynamic range, 16-bit precision and 125 MHz sampling rate. Choosing such a high-speed ADC chip makes it possible to compare the performance of curve fitting and the neural network model with different number of sampling points.

To prepare the datasets, we watch two channels of signals simultaneously. One channel is the trigger signal driving the LED, and the other channel is the output of the shaper on the FEE card. We randomly choose a fixed-interval section from the most salient part of the output pulse. Then we add a label to the pulse according to the interval between the trigger signal and the selected section. This label is used to train the neural network and work as the baseline for curve fitting. We normalize the amplitude of the ADC sampling points to the range similar to section \ref{sec:curve-fitting-sim} and section \ref{sec:nn-sim}. We collect 80000 samples for the training dataset and 20000 samples for the test dataset.

\subsection{1 \texorpdfstring{$\mu$}{u}s shaping time}
\label{sec:exp-1us}

\begin{figure}[htbp]
	\centering
	\subfigure[timing resolution]{			
		\includegraphics[width=0.48\textwidth]{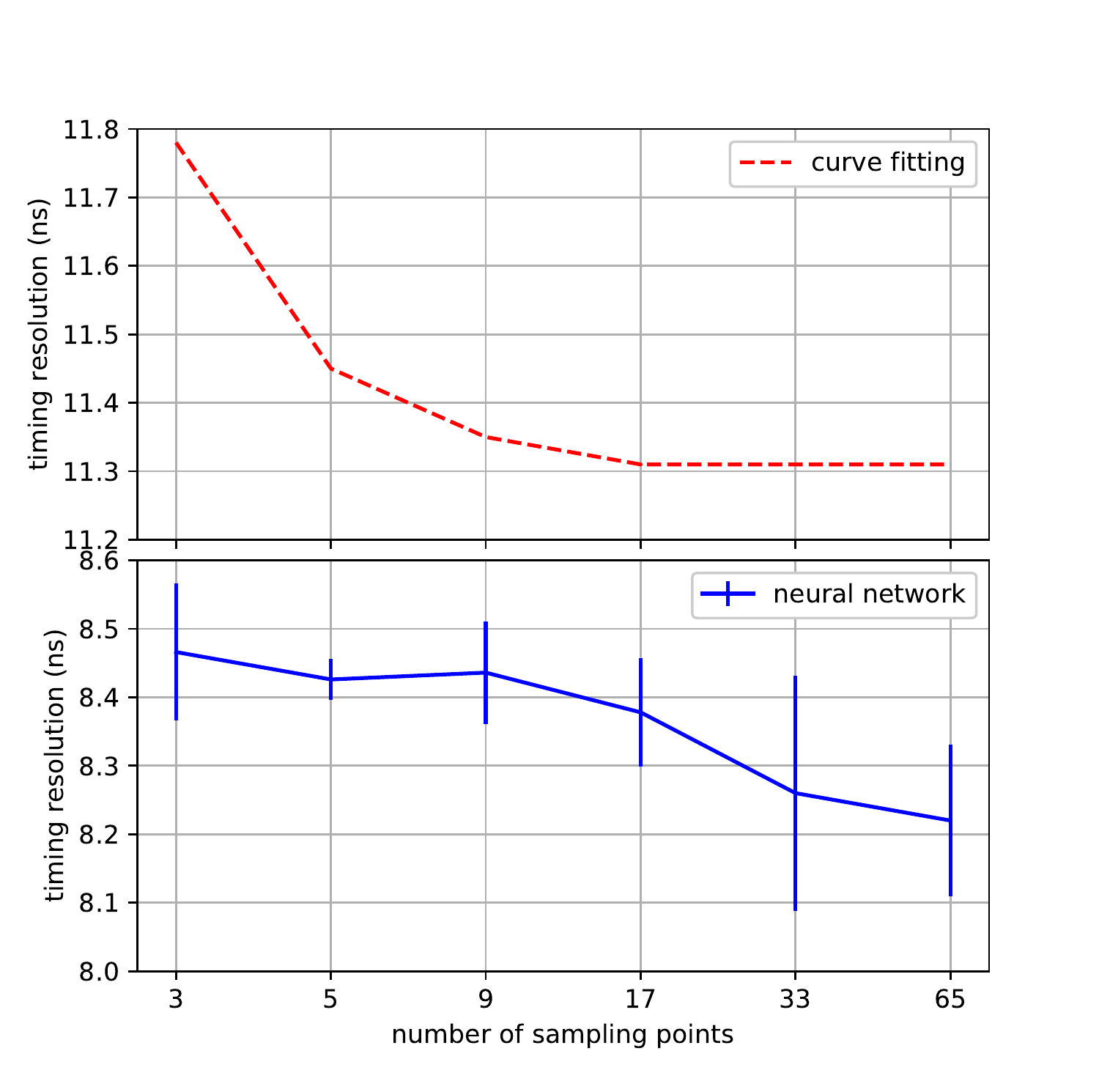}}		
	\subfigure[system bias]{
		\includegraphics[width=0.48\textwidth]{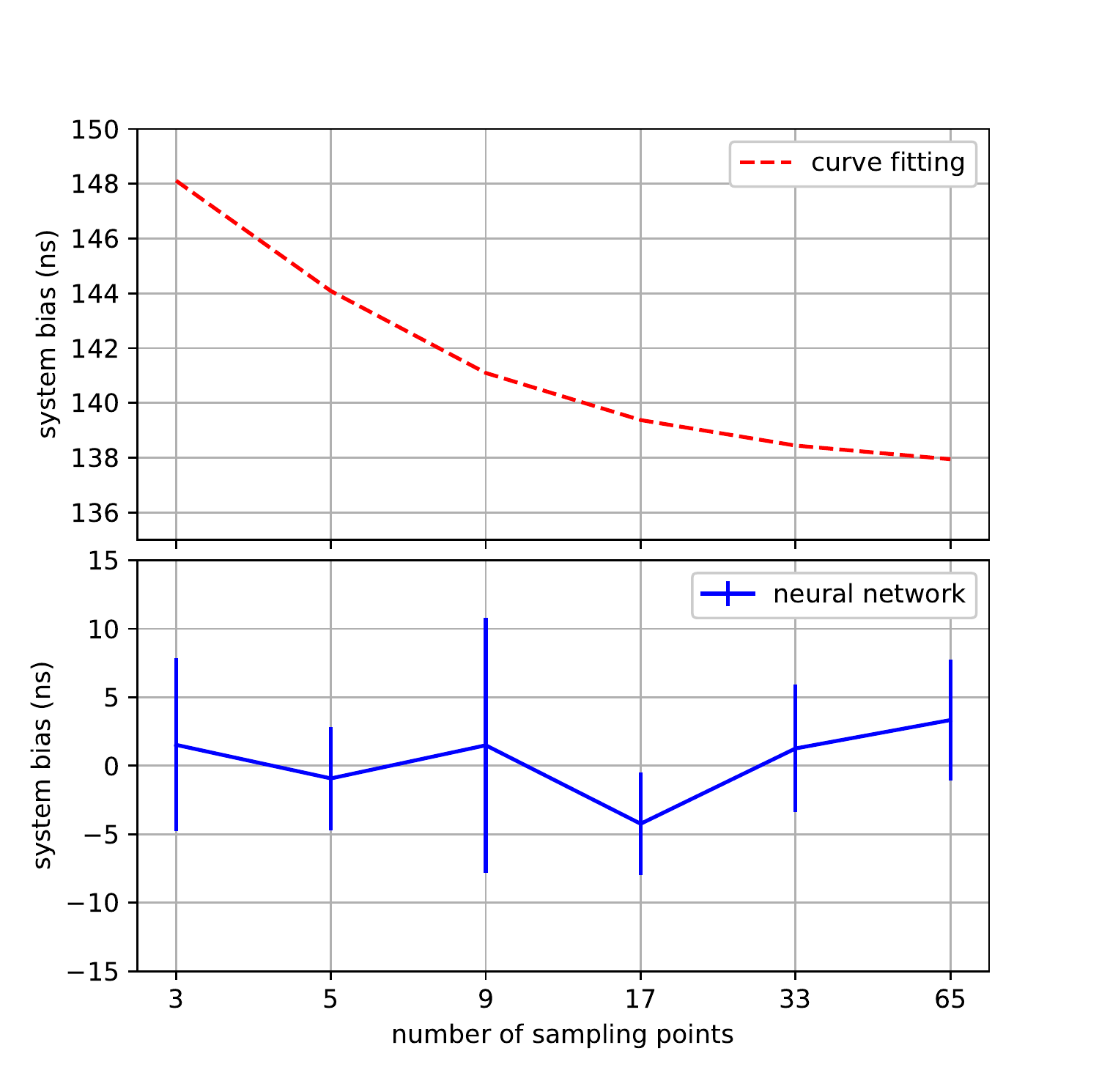}}
	\caption{\label{fig:exp-1u} Experimental results for the 1 $\mu$s shaping time.}
\end{figure}

In this part, we conduct experiments with 1 $\mu$s shaping time (2 $\mu$s peaking time) which is the ALICE PHOS specification. The sampling section has a span of 3072 ns. We choose $2^k + 1$ points evenly distributed in the sampling section. These points have been stretch to a fixed length of 256 points by cubic (for $k \geq 2$) or quadratic (for $k = 1$) interpolation when training the neural network. We pre-train the model under the assumption of the Gaussian noise with the parameterization in section \ref{sec:nn-sim}. Then we finetune the whole network using the experimental data. The base network is trainable when finetuning.

We analyze 6 different conditions with $k = 1, 2, 3, 4, 5, 6$. This gives an approximate sampling rate of 0.625 MHz, 1.25 MHz, 2.5 MHz, 5 MHz, 10 MHz and 20 MHz respectively. We perform an independent training process using the same pre-trained model. Then we test our model on the corresponding test dataset and make a Gaussian fit of the residuals (difference between regression outputs and time labels) to get the mean and the standard deviation. The standard deviation of the Gaussian fit is a measure of the timing resolution and the mean is a measure of the system bias. For curve fitting, we use the same sampling points and fit the residuals (difference between fitting parameters and time labels) to a Gaussian distribution.

We use a batch size of 16 when training the neural network, and the training proceeds for 10 epoches. The final result and error bar ($1\sigma$ error) for the neural network are calculated by the test results paused at even number of training epoches.

The main result is shown in figure \ref{fig:exp-1u}. In the left figure, it can be seen that the neural network works better than curve fitting steadily. With as few as 3 sampling points, the two methods can already achieve relatively good performance. When sampling points increase, the results improve slightly. When we use greater or equal than 17 sampling points, the performance of curve fitting hits a plateau, but the neural network can still improve. The best performance achieved by the neural network is $8.22\pm0.11$ ns, which is $27.3\%$ better than curve fitting (11.31 ns).

In the right figure, the system bias is greatly reduced by the neural network model compared to curve fitting. From directly observation, the interval between the start of the trigger signal and the start of the shaped pulse is approximately 15 sampling points (120 ns), which is close to results from curve fitting (137.94 ns to 148.11 ns). The bias for the neural network fluctuates around the horizontal axis. Since the bias is a fixed value for a given model, it can be calibrated in the same way as curve fitting, and the burden for calibration is considerably alleviated.

\subsection{100 ns shaping time}
\label{sec:exp-100ns}

\begin{figure}[htbp]
	\centering
	\subfigure[timing resolution]{
		\includegraphics[width=0.48\textwidth]{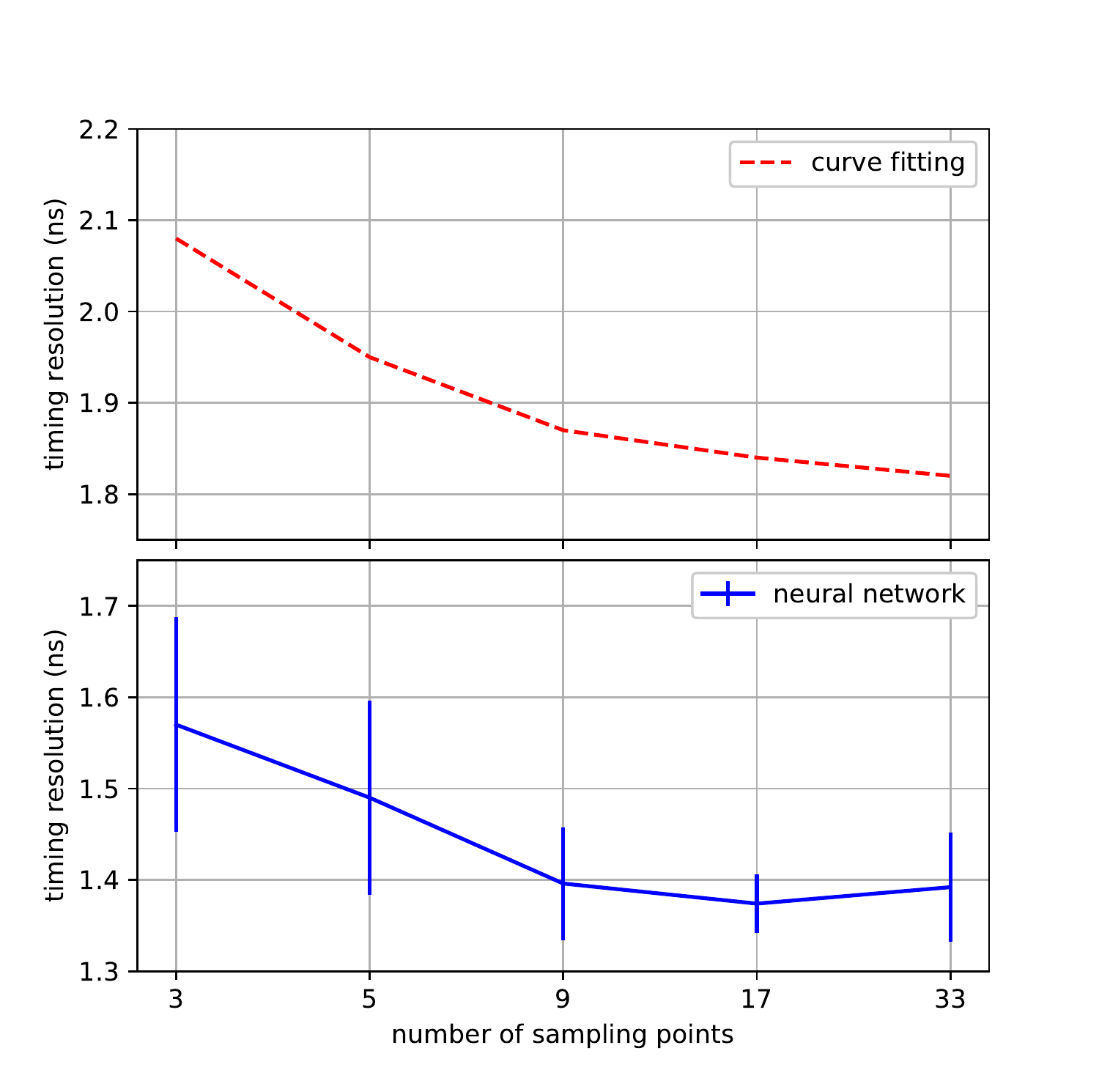}}		
	\subfigure[system bias]{
		\includegraphics[width=0.48\textwidth]{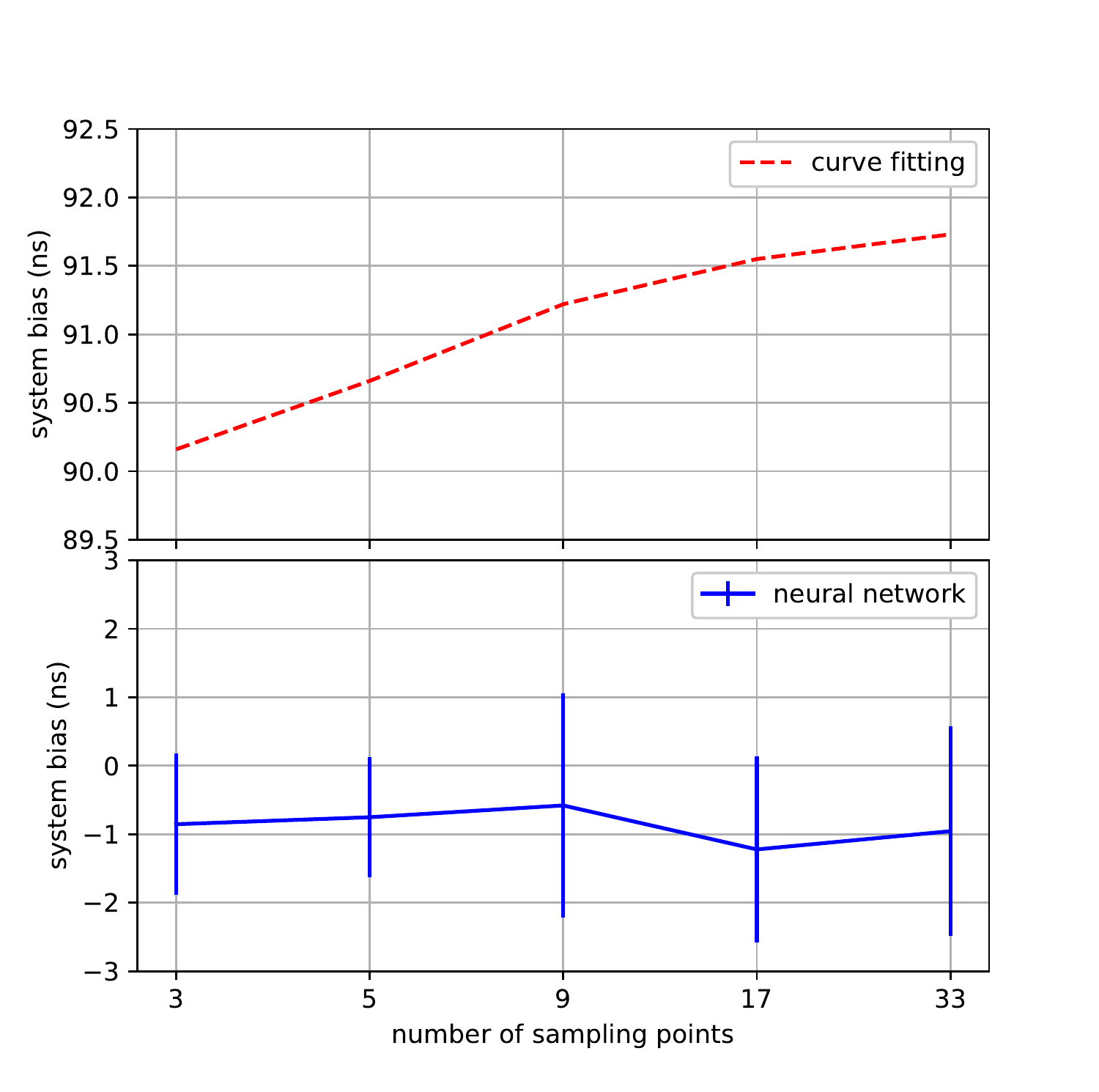}}
	\caption{\label{fig:exp-100n} Experimental results for the 100 ns shaping time.}
\end{figure}

In this part, we conduct experiments with 100 ns shaping time (200 ns peaking time) which is the ALICE EMCal specification. We replace resistors and capacitors in the CR-RC2 shaper on the FEE card to achieve a shorter shaping time. The sampling section has a span of 256 ns. We choose $2^k + 1$ points and analyze 5 different conditions with $k = 1, 2, 3, 4, 5$. This gives a sampling rate of 7.8125 MHz, 15.625 MHz, 31.25 MHz, 62.5 MHz and 125 MHz respectively. Other experimental conditions and procedures are similar to section \ref{sec:exp-1us}.

To determine the label for curve fitting and the neural network with a precision superior to the sampling period, we fit the trigger signal to the square pulse response of a second-order system:

\begin{gather}
	Y_{\text{step}}(t) = K\left(1 + \frac{T_1}{T_2 - T_1}e^{-(t-t_s)/T_2} - \frac{T_2}{T_2 - T_1}e^{-(t-t_s)/T_1}\right)u(t-t_s) \\
	Y_{\text{square}}(t) = Y_{\text{step}}(t) - Y_{\text{step}}(t-w)
\end{gather}

\noindent where $u(t)$ is the step function, $Y_{\text{step}}(t)$ is the overdamped step response of a second-order system. $K$ and $t_s$ are parameters to be fitted, and other parameters are fixed according to the circuit specification and the experimental observation. $t_s$ is used as the label to judge the quality of curve fitting and train the neural network.

The main result is shown in figure \ref{fig:exp-100n}. In the left figure, the timing resolution has improved significantly compared to the 1 $\mu$s shaping time. Again, the neural network outperforms curve fitting. When the number of sampling points increases from 3 to 33, the precision of the neural network and curve fitting increases slightly, and the trend gradually slows down. The neural network achieves the optimal result $1.37\pm0.03$ ns at 17 sampling points, which is $24.7\%$ better than curve fitting (1.82 ns).

In the right figure, the system bias of the neural network model is much less than curve fitting. Basically, curve fitting has a large system bias (90.16 ns to 91.73 ns) which is in accord with approximate 96 ns from direct observation, but the neural network model suppresses the absolute value of the bias to less than 2 ns. This facilitates the calibration and improves the overall stability of the timing system.

\subsection{Discussion about the experimental results}

In the above experiments, a relation between the shaping time of the shaper and the timing resolution is being considered. The experimental results show that, decreasing shaping time can potentially increase the timing resolution when other conditions are the same. In the frequency domain, a shorter shaping time means a bandpass filter with higher cut-off frequency. Therefore, more information about the original event is kept. In the time domain, shorter shaping time can alleviate the long-range misfit problem. To be more specific, in the experiments of 1 $\mu$s shaping time, sampling points are far away from the desired start time $t_0$; thus any slight discrepancy between the fitted model and the ideal model will cause a large deviation in the value of $t_0$. The similar issue applies to the neural network if we view the discrepancy as an intrinsic error and a source of misunderstanding. To use the 100 ns shaping time, the distance between sampling points and the start time is shortened and the long-range problem is properly handled.

However, on the other hand, when the shorter shaping time is used, the influence of three kinds of variations (especially short-term change and random noise) is relatively more significant. Besides, since the width of the LED pulse is less than 50 ns, signal integrity issues (especially overshooting) affect the precision of the fitted label. As a result, the improvement of timing resolution is worse than estimates based on a proportional hypothesis ($\sim$0.8 ns). If we use auxiliary timing detectors to construct a coincidence measuring system, better results can be expected.

\section{Conclusion}
\label{sec:conclusion}

The classic curve fitting method uses a Gaussian noise hypothesis, and its performance is guaranteed by its statistical properties. However, when the long-term drift, short-term change and random noise are presented in the pulse function, the limitation of curve fitting emerges. Among the possible alternatives, neural networks show strong resistance to these three kinds of variations by its delicate structure and optimization process. Simulations and experiments demonstrate its superiority over curve fitting.

Nevertheless, neural networks have their special requirements which pose new challenges to the design of the detector system. Since most deep learning methods are based on the supervised learning, an accurate label for training is needed. Sometimes acquiring the label is not an easy task, especially when the detector system has complex geometric structures and intricate components. This raises the demand for the traceable design, i.e. a design scheme in which the timing information can be traced back internally through the calibration process. From this perspective, we sincerely hope our work will provide a new way of thinking in the future design of timing systems.

\acknowledgments

This research is supported by the National Natural Science Foundation of China (Grant Number 11875146, 11505074, 11605051).

\input{pulse_manu.bbl}

\end{document}

%% file: pulse_manu.bbl
\providecommand{\href}[2]{#2}\begingroup\raggedright\endgroup